\documentclass[12pt,preprint]{aastex}
\usepackage{lscape}
\usepackage[latin1]{inputenc}
\usepackage{amsmath}
\usepackage{amsfonts}
\usepackage{color}
\usepackage{epsfig}
\usepackage{amssymb, amsmath}
\usepackage{graphicx,natbib}
\usepackage{multirow}
\usepackage{float}
\newcommand{\uma}[1]{^{\mathrm{#1}}}
\newcommand{\dma}[1]{_{\mathrm{#1}}}
\newcommand{\amin}{a\dma{min}}

\small

\slugcomment{}

\shorttitle{}
\shortauthors{B. Mennesson et al.}

\defcitealias{Absil09}{Paper~I}

\def\deg{\hbox{$^\circ$}}                  
\def\fdeg{\hbox{$.\!\!^\circ$}}            

\begin{document}


\title{An interferometric study of the Fomalhaut inner debris disk\\
II. Keck Nuller mid-infrared observations }


{\author{B. Mennesson\altaffilmark{1}, O. Absil\altaffilmark{2}, J. Lebreton\altaffilmark{3}, J.-C. Augereau\altaffilmark{3}, E. Serabyn\altaffilmark{1}, M.M. Colavita\altaffilmark{1}, R. Millan-Gabet\altaffilmark{4}, W. Liu\altaffilmark{5}, P.Hinz\altaffilmark{6}, P. Th\'{e}bault\altaffilmark{7}}





\altaffiltext{1}{Jet Propulsion Laboratory, California Institute of Technology, 4800 Oak Grove Drive, Pasadena CA 91109-8099, USA}
\altaffiltext{2}{D\'{e}partement d'Astrophysique, G\'{e}ophysique et Oc\'{e}anographie, Universit\'{e} de Li\`{e}ge, 17 All\'{e}e du Six Ao\^{u}t, B-4000 Sart Tilman, Belgium }
\altaffiltext{3}{IPAG, UMR 5274, CNRS and Universit\'{e} Joseph Fourier, BP 53, F-38041 Grenoble, France}
\altaffiltext{4}{Michelson Science Center, California Institute of Technology, 770 South Wilson Avenue, Pasadena CA 91125, USA}
\altaffiltext{5}{Infrared Processing and Analysis Center, California Institute
of Technology, Mail Code 100-22, Pasadena, CA 91125, USA}
\altaffiltext{6}{Steward Observatory, University of Arizona, 933 N Cherry Avenue,
Tucson, AZ 85721, USA}
\altaffiltext{7}{Observatoire de Paris, Section de Meudon, F-92195 Meudon Principal Cedex, France}


\begin{abstract}

We report on high contrast mid-infrared observations of Fomalhaut obtained with the Keck Interferometer Nuller (KIN) showing a small resolved excess over the level expected from the stellar photosphere.  The measured null excess has a mean value of 0.35\% $\pm$ 0.10\% between 8 and 11 $\mu$m and increases from 8 to 13 microns.  Given the small field of view of the instrument, the source of this marginal excess must be contained within 2AU of Fomalhaut. This result is reminiscent of previous VLTI  K-band ($\simeq$ 2$\mu$m) observations, which implied the presence of a $\sim$ 0.88\% excess, and argued that thermal emission from hot dusty 
grains located within 6 AU from Fomalhaut was the most plausible explanation. 
Using a parametric 2D radiative transfer code and a Bayesian analysis, we examine different dust disk structures to reproduce  both the near and mid-infrared data simultaneously. While not a definitive explanation of the hot excess of Fomalhaut, our model suggests that the most likely inner few AU disk geometry consists of a two-component structure, with two different and spatially distinct grain populations. The 2 to 11 microns data are consistent with an inner  hot ring of very small ($\simeq$ 10 to 300 nm) carbon-rich grains concentrating around 0.1\,AU. The second dust population - inferred from the KIN  data  at longer mid infrared wavelengths - consists of larger grains (size of a few microns to a few tens of microns) located further out in a colder region where regular astronomical silicates could survive, with an inner edge around 0.4\,AU to 1\,AU. 
From a dynamical point of view, the presence of the inner concentration of sub-micron sized grains is surprising, as such grains should be
expelled from the inner planetary system by radiation pressure within only a few years. This could either point to some inordinate replenishment rates (e.g. many grazing comets coming from an outer reservoir) or to the existence of some braking mechanism preventing the grains from moving out.

\end{abstract}






\section{Introduction}

Fomalhaut is a bright ($V=1.2$) relatively young ($\sim$ 200 Myr, \citet{Difolco04}) A4 main sequence star located 7.7\,pc away featuring a large far infrared (IR) excess first detected by IRAS \citep{Aumann85}. The suspected surrounding debris disk was first resolved in the far-IR by the Kuiper Airborne Observatory \citep{Harvey96}, then in the sub-millimeter \citep{Holland98}, and more recently in scattered light using HST/ACS \citep{Kalas05}. The latter optical observations show a belt of cold dust concentrated around 140 AU from the star, with an asymmetric structure consistent with gravitational sculpting by an orbiting planet \citep{Kalas05,Quillen06}. Recent ALMA 350 GHz observations of the Fomalhaut debris ring \citep{Boley12} demonstrate that the disk parent body population is 13 to 19 AU wide with sharp inner and outer boundaries, and also suggest that debris confined by shepherd planets is the most likely origin for the ring's observed morphology. Interestingly, the detection of a point source located at 118 AU from the star - just inside the inner edge of the ring, matching its predicted location - was reported in 2008  using HST's Advanced Camera for Surveys (ACS) \citep{Kalas08}.  It was interpreted as a direct image of the disk-perturbing planet, and named Fomalhaut b. However this detection was made at ~600-800 nm, and no corresponding signatures have been found so far in the near-IR range \citep{Kalas08, Marengo09} , where the bulk emission of such a planet should be expected, nor at 4.5 $\mu$m with Spitzer / IRAC \citep{Janson12}. The true nature of this HST detected source hence remains unclear at this point, but all current models involve dust. It could either be in the form of a large circumplanetary disk around a massive planet, or created via a recent collision between two Kuiper Belt-like objects of radii about 50 km \citep{Currie12,Galicher12}.

Far inside this ring, within a few tens of AUs from the star, a warm dust component has also been detected by Spitzer \citep{Stapelfeldt04}: the InfraRed Spectrograph (IRS) measured an excess continuously increasing with wavelength between 17.5 and 34 microns, while direct images obtained with the MIPS instrument (Multiband Imaging Photometer for Spitzer) resolved the circumstellar disk down to a wavelength of 24 microns. Both of these observations point to a region of compact residual excess emission extending inward of $\sim$ 20 AU, but whose spatial and physical structure can not be uniquely determined given the limited resolution of Spitzer. The presence of warm dust close to the star is independently confirmed by recent 70 $\mu$m images of Fomalhaut obtained with Herschel/ PACS (Photodetector Array camera and Spectrometer) \citep{Acke12}.  These spectacular images show both the cold outer ring and an unresolved excess source co-located with the central star (i.e within 5.7", given Herschel beam size) carrying 50\% $\pm$ 10\% of its flux. 


High accuracy long baseline interferometric observations can provide the resolution and contrast required  to probe this central region in greater detail. They can focus on the very inner part of Fomalhaut's environment (within a few AU), probing very different astrophysical scales and conditions than those studied by single optical/ IR telescopes.  near-IR interferometric observations obtained at the VLTI were already  reported in a first paper \citep[][hereafter \citetalias{Absil09}]{Absil09}, concluding that a K-band excess of $0.88 \pm 0.12$\% is present, and arguing that thermal emission from hot dust grains located within 6\,AU of Fomalhaut is the most plausible explanation for the detected excess. We present here new data obtained at N band (8 to 13\,$\mu$m) with the Keck Interferometer Nuller (KIN) in August 2007 as part of KIN's commissioning ``shared risk'' science operations, and in July 2008 as part of the KIN key science program (PI: Phil Hinz). These near-IR (VLTI) and mid-IR (Keck) interferometric measurements are then used in conjunction with spectro-photometric data to constrain the physical parameters of the inner few AU of Fomalhaut's debris disk. Finally, we discuss a few possible astrophysical scenarios which are compatible with the observed disk and dust characteristics.


\section{Nulling Set-Up and Observable Quantities}

The overall KIN system architecture and performance are presented in full detail in recent publications \citep{Colavita08,Colavita09,Serabyn2012}.  In brief, four beams are recombined by the KIN system. A split mirror located downstream of each Keck telescope adaptive optics system---close to a pupil plane---divides the light gathered by each telescope into ``left'' and ``right'' beams. Interferometric nulling occurs separately between the two Keck left beams, and between the two right beams. The resulting nulled output fields are then coherently recombined using a standard Michelson interferometer, called the ``cross-combiner".  As the optical delay is rapidly scanned inside the cross-combiner, one first measures the cross combiner fringe amplitude at null in each of ten independent spectral channels covering the full N band (8 to 13\,$\mu$m), and then "at peak". The null depth is defined as the ratio of the cross-combiner fringe amplitudes obtained at null and at peak.  The rationale for this complex 4-beam combination  and modulation, is that the resulting measured null depth is free of slow drifts in the incoherent background, a source of strong potential bias for  ground-based interferometric observations in the thermal IR. Two different scales and baselines are then involved: the interferometric nulling baseline of length $B \simeq 85$\,m, separating the telescopes centers, and the short cross combiner baseline $b \simeq 4$\,m,  characteristic of the interference between the ``left'' and ``right'' parts of a given Keck telescope.

For a perfect instrument, defined as providing a null depth of zero on a point source, one can relate the measured monochromatic astrophysical null $N_{\rm ast}$ to the source brightness distribution on the sky $I(\vec{\theta})$. The  observed null can be expressed in the following way \citep{Serabyn2012} 

\begin{equation}
N_{\rm ast}(\lambda) = \frac{\int I(\vec{\theta}) \sin^{2}(\pi\vec{B} \cdot \vec{\theta}/\lambda)) \cos(2\pi\vec{b} \cdot \vec{\theta}/\lambda) \sqrt{T_L(\vec{\theta})T_R(\vec{\theta})} d\vec{\theta}} {\int I(\vec{\theta}) \cos^{2}(\pi\vec{B} \cdot \vec{\theta}/\lambda) \cos(2\pi\vec{b} \cdot \vec{\theta}/\lambda)) \sqrt{T_L(\vec{\theta})T_R(\vec{\theta})} d\vec{\theta}} \; , \label{eq:nulldef}
\end{equation}
where $T_L(\vec{\theta})$ and $T_R(\vec{\theta})$ designate the sky transmission patterns of the left and right Keck beams, respectively. They are computed from the telescopes orientation and from the overall beam train propagation, which includes an intermediary focal plane pinhole. As a result, the field of view FWHM is at maximum 450\,mas, along the direction perpendicular to the left-right split, corresponding to about $\pm$ 2AU at Fomalhaut's distance.
As shown in Equation~\ref{eq:nulldef}, in the case of an extended source, the measured null level is not only affected by the long baseline nulling pattern (fast oscillating squared sine term), but also by the cross fringe pattern (slowly oscillating cosine term) \footnote{A curious effect of the KIN 4-beam combination is that for emission sources extending further out than $\lambda/b$ in the direction of the cross-combiner baseline b (b $\simeq 4\,m$), some regions will contribute "negatively", i.e. effectively decrease the observed null depth. This is illustrated by the areas of negative transmission shown in Fig.~\ref{fig:tmap12}} and by the lobe antenna of each single beam. As representative examples of these 3 contributors to the effective KIN sky transmission, the left and right panels of Fig.~\ref{fig:tmap12} show the monochromatic (10\,$\mu$m) KIN's transmission patterns corresponding to the 2007 and 2008 observations of Fomalhaut, respectively. Both are derived at the time of meridian transit and mostly differ in the orientation of the short baseline (see section 3).
Also indicated in these figures is the orientation of Fomalhaut's outer debris disk major axis ($156\deg$ East of North), as imaged by HST/ACS \citep{Kalas05}. For both epochs, the projected Keck to Keck interferometric baseline orientation is $\simeq$ 73 degrees away from the outer dust disk main axis. 


In the case where no extended emission is present (source angular size $\ll \lambda/b$), 
the astrophysical null expression (Equation~\ref{eq:nulldef}) can be approximated by:
\begin{equation}
N_{\rm ast}(\lambda)=\frac{\int I(\vec{\theta}) \sin^{2}(\pi\vec{B} \cdot \vec{\theta}/\lambda) d\vec{\theta}} {\int I(\vec{\theta}) d\vec{\theta}} \; . \label{eq:nullapprox}
\end{equation}
In particular, for a naked star represented by a uniform disk (UD) of diameter $\theta_{\ast}$ ($\ll \lambda/B$), the observed astrophysical null is given by:
\begin{equation}
N_{\rm ast}(\lambda) = \left(\frac{\pi B\theta_{\ast}}{4\lambda}\right)^2 \; . \label{eq:nullUD}
\end{equation}

For a more realistic model, in which a naked star is represented by a limb darkened disk of diameter $\theta_{\rm LD}$, with a linear limb darkening coefficient $u_{\lambda}$, the observed astrophysical null is \citep{Absil06,Absil11}:
\begin{equation}
N_{\rm ast}(\lambda) = \left(\frac{\pi B\theta_{\rm LD}}{4\lambda}\right)^2 \, \left(1-\frac{7u_{\lambda}}{15}\right)   \, \left(1-\frac{u_{\lambda}}{3}\right)^{-1} \ \; . \label{eq:nullLD}
\end{equation}

In practice, the null depth measured on a point source is not zero but equal to the instrumental null, noted $N_{\rm ins}$. Consequently, the observed null depth $N_{\rm obs}$ is not strictly equal to the astrophysical null $N_{\rm ast}$ characteristic of the source, and one measures instead:
\begin{equation}
N_{\rm obs}=(N_{\rm ast}+N_{\rm ins})/(1+N_{\rm ast}N_{\rm ins}) \; .
\end{equation}
Full details on the terms contributing to the instrumental null can be found in \citet{Colavita09} and \citet{Serabyn2012}. 
For all the observations considered here, both $N_{\rm ins}$ and $N_{\rm ast}$ are small ($<2\%$), so that one can use the approximation:
\begin{equation}
N_{\rm obs}=N_{\rm ast}+N_{\rm ins} \; . \label{eq:nullcalib}
\end{equation}

As in classical stellar interferometry, the instrumental null  $N_{\rm ins}$ is derived from nulling observations of calibrator stars with known diameters and limb darkening properties, i.e., with predictable astrophysical nulls (Equation~\ref{eq:nullLD}).


\section{Observations and Data Analysis}

The basic observing block of the KIN is a 400ms long  "null/ peak micro-sequence" (see \citet{Colavita09}, figure 3). Each micro-sequence yields an individual null depth estimate defined as the ratio of the cross-combiner fringe amplitudes measured at null and at peak.  A null measurement sequence consists in 1000 consecutive micro-sequences, from which a mean null estimate is derived, together with its standard deviation.   

Six such null measurement sequences were recorded on Fomalhaut in 2007: one on August 28 and five more on August 30. Only the six spectral channels with wavelengths shorter than 11\,$\mu$m provided adequate signal to noise and are discussed here. Eight more individual null measurement sequences were then obtained on July 17 and 18, 2008. This time, the full 8 to 13\,$\mu$m range was covered using ten spectral channels.  

\subsection{Calibrators}

Fomalhaut's observations were interleaved with measurements of four nearby calibrator stars, whose characteristics are given in Table~\ref{tab:cal}. Their limb darkened (LD) diameters are estimated from V and K magnitudes (corrected for interstellar reddening) and surface brightness relationships developed by  \citet{diBenedetto2005}, except for the redder source HD 214966.  In this case, the V-K color falls outside of Di Benedetto's relationships applicability range, and  we use instead the surface brightness relationships derived by \citet{Bonneau}.


At the central wavelength of each spectral channel, the expected calibrators astrophysical nulls are then derived using Equation~\ref{eq:nullLD}, assuming a constant N-band linear limb darkening coefficient $u_N=0.12$ \citep{Tango02} for all four calibrators. Limb darkening corrections are expected to be small for such stars at 10\,$\mu$m, and fairly constant over the range of $T_{\rm eff}$ and $\log(g)$ covered by our calibrators. 

While Di Benedetto (2012) reports relative uncertainties as low as  2\% in his LD diameter estimations, we use larger (1-$\sigma$) error bars to account for uncertainties in K band photometry or mid-IR LD coefficients.  For HD 222547 and HD215167, which have accurate K band photometry, we adopt a relative error of 6\%. For HD 214966, we use the 7\% relative error quoted by \citet{Bonneau}. Finally, HD 210066 only has 2MASS (saturated) photometric measurements available at K-band, and we adopt a conservative LD diameter error of 10\%. 

\subsection{Calibrated Astrophysical Nulls}

The instrumental null at the time of a given Fomalhaut observation is computed by interpolation between the instrumental nulls derived on adjacent calibrators. At each wavelength and for each projected baseline length, one computes Fomalhaut's calibrated astrophysical null from the observed nulls using Equation~\ref{eq:nullcalib}. The resulting calibrated astrophysical nulls are summarized in Table~\ref{tab:obs}.

Overall, the length of the long baseline $\vec{B}$ projected onto the sky plane ($B_p$) varied between 55.12\,m and 77.81\,m during the observations, while the projected baseline azimuth varied only slightly (between $47\fdeg4$ and $49\fdeg9$). 

The projected length of the short baseline $\vec{b}$, on the other hand, does not change: by construction, it is always located in the (pupil) plane perpendicular to the line-of-sight (see Fig.~\ref{fig:config}).  We note that the short baseline orientation was changed by $90\deg$ between the 2007 and 2008 observations. In 2007, it was located in the plane defined by the zenith and the line-of-sight, while in 2008, it was located in the local horizontal plane (see Fig.~\ref{fig:config}). These variations in the short baseline azimuth do not have a large influence on the measurements but were included in the null computations (Equation~\ref{eq:nulldef}) and modeling. 


\subsection{Null excess leakage}

The next step in the data analysis is to evaluate the fraction of Fomalhaut's calibrated astrophysical null that actually comes from the circumstellar environment. This requires the computation of the null depth expected from the photosphere alone (naked star scenario, Equation~\ref{eq:nullLD}). We use Fomalhaut's limb darkened diameter $\theta_{\rm LD} = 2.223 \pm 0.022$\,mas deduced from the latest VLTI K band measurements (\citetalias{Absil09}) and a constant linear limb darkening coefficient $u_N=0.06$ between 8 and 13\,$\mu$m \citep{Tango02}. For the baselines considered here, the stellar contribution to the total null depth typically ranges from 0.5\% and 0.2\% for wavelengths from 8 to 13\,$\mu$m. The difference between Fomalhaut's observed calibrated null and this purely photospheric leakage is noted "null excess'' hereafter. It characterizes the contribution of any source of circumstellar emission located within the nuller field of view.   

For each of the 14 independent calibrated observations reported in Table~\ref{tab:obs}, we compute the resulting null excess as a function of wavelength. Since there is very little variation in azimuth over the full data set but some projected baseline length variations and some changes in the instrumental set-up between 2007 and 2008, we further regroup the measurements according to mean baseline length (`"short'' or "long'') and year of observation. This allows us to generate 4 final data sets from the original 14, which reduces null measurements uncertainties and the modeling computation time (section 4). For instance the 2007 "short'' baseline data summarize measurements from the three shortest baselines, ranging from 58 to 63\,m. From that sub-ensemble of measured null excesses and associated error bars, we compute the weighted mean null excess and its 1-$\sigma$ uncertainty (we use the weighted standard deviation). This provides an accurate estimate of the null excess at the 2007 short observing baseline, which has a mean length of 60.72\,m and an equivalent azimuth of $49.77\deg$. Fig.~\ref{fig:null78} (top left) shows Fomalhaut's measured null excess at this baseline, as a function of wavelength.  Fig.~\ref{fig:null78} (bottom left) shows similar curves for the 2007 long baseline data (mean length of 71.88\,m, for an equivalent azimuth of $48.66\deg$).  Analogous results are presented for the 2008 observations in Fig.~\ref{fig:null78} (right panels), grouping the four shortest baselines (55 to 62\,m, mean length of 58.43\,m, azimuth of $49.42\deg$, top right inset), or the four longest ones (73 to 78\,m, mean length of 75.79\,m, azimuth of $47.40\deg$, bottom right inset). These figures show that our 2007 and 2008 data sets are consistent with each other (generally to within 1$\sigma$), although the short wavelengths excess null looks slightly larger in 2007 than in 2008.






Based on the results presented on Fig.~\ref{fig:null78} and taking all 2007 and 2008  observations into account, the (weighted) average null excess measured between 8 and 11\,$\mu$m is $0.35\%$. 
Small systematic errors have been previously identified in the KIN data \citep{Colavita09}, potentially biasing the nulls measured at different wavelengths in a systematic way during a given night. As a consequence, we use for our measurements the 1-$\sigma$ uncertainty level recommended in \citet{Colavita09}, which represents our best understanding of the instrument. For KIN measurements of stars as bright as Fomalhaut in the 8 to 11\,$\mu$m range, this uncertainty  corresponds to 0.2\% rms \textit{per night}, or 0.1\% over the 4 nights of observations reported here. Our final estimate of Fomalhaut's averaged null excess in the 8 to 11 $\mu$m region is then $0.35\% \pm 0.10\%$. Finally, we note that for the 2008 data - covering the whole 8 to 13 $\mu$m atmospheric N band window-, the observed excess increases with wavelength, for both baseline ranges considered.   






\section{Modeling and Interpretation}

Although each of our four data subsets is formally consistent with  pure photospheric emission to within 2-3 $\sigma$, an excess null is measured at all wavelengths and for {\it all four}  of the observing nights, covering years with slightly different KIN instrumental set-ups. This suggest the need to go beyond a simple photospheric model to explain the observations. Owing to the various observational evidence for warm emission in the Fomalhaut inner system (\citetalias{Absil09}, \citet{Stapelfeldt04, Acke12}), we hereafter model the measured excess null using purely morphological debris disk models. In particular, we do not favor the assumption that either the observed KIN or the VLTI /VINCI emission is due to gaseous free-free
emission from a stellar wind, and there are three main reasons for that: 
\begin{itemize}
\item For A stars, the photospheric emission still goes as $\nu^{2}$ in the near-IR, while the $\nu ^{0.6}$ spectral slope for free-free emission starts to break down in the 
mid-IR, where free-free emission becomes optically thin, and the 
slope flattens for shorter wavelengths (e.g. \citet{wb75}). As a result, the free-free emission level relative to the star is expected to be smaller in the near-IR than in the mid-IR.  Free-free emission would probably not be sufficient to reproduce the 
observed VINCI K-band excess (\citetalias{Absil09}). And if it were, then it would be stronger in the mid-IR, which is not compatible with the low level of emission seen by the KIN.
\item Second, the size of free-free emission regions decrease at 
shorter wavelengths. Actually, at the wavelength where free-free emission 
becomes optically thin, the size of the emitting region is of the order of the 
stellar size. Therefore, we would not expect the near-IR free-free 
emission to be significantly extended (while the previous VLTI observations show the K-band 
emitting region to be at least 10 times larger than the photosphere).
\item Third, stellar models predict very small mass loss rates for A-stars (e.g., $10^{-16}$ 
$M_{\odot}$/yr, \citet{Babel95}), which are not compatible with strong free-free emission.  
\end{itemize}

\subsection{Modeling the Mid-Infrared Data with a Solar Zodiacal Disk Model}

Since the observed null depth is a function of source brightness distribution and baseline orientation (see Equation~\ref{eq:nulldef}), we first need to define a morphological model for any excess emission before computing its associated photometric flux. In other words, the true astrophysical excess can not be uniquely derived from the data, unless some assumptions are made on its spatial brightness distribution. As a first rough attempt to estimate the luminosity of the marginal excess detected, we have used a scaled model of the solar system's zodiacal disk around Fomalhaut.  This model is based on the parametric description of the zodiacal cloud observations by the COBE/DIRBE instrument \citet{Kelsall98} and is implemented in the \textsc{Zodipic} package\footnote{\textsc{Zodipic} is an IDL program for synthesizing images of exozodiacal clouds that can be downloaded at {\tt http://asd.gsfc.nasa.gov/Marc.Kuchner/home.html}.}. The dust density and temperature profiles are assumed to follow the power laws derived in the solar system case: $ n(r) \propto r^{-1.34}$, and $T(r) \propto L^{\delta/2}/r^{\delta}$, with $\delta$=0.467 \citep{Kelsall98} and Fomalhaut's luminosity $L= 17.7 \,L\sun$ \citep{Difolco04}. The dust is assumed to extend from the sublimation radius corresponding to a temperature of $1500\,$K (typical of silicate grains found in the solar system), all the way to 10 AU, i.e. much further out than the KIN field of view. The exo-zodi disk is modeled with the same inclination (65.6$\deg$) and position angle (156$\deg$) as Fomalhaut's outer dust disk \citep{Kalas05}. 
The result of the best fit is a luminosity (and density) scaling factor of 350 with respect to the solar case, with 1-$\sigma$ uncertainty of about 100. Such a disk would produce a total emission of 0.55\,Jy across the KIN field-of-view at 10\,$\mu$m, i.e., a flux ratio of  $\simeq$ 3\% with respect to the photosphere. This illustrates the fact that due to the nuller sky transmission pattern, the true 10 $\mu$m astrophysical excess around Fomalhaut could be significantly larger than our observed null excess of $\simeq$ 0.35\%. This is especially true in the case of Fomalhaut, where the disk PA is almost perpendicular to the long Keck to Keck baseline (Fig.~\ref{fig:tmap12}).   

While such a simple zodiacal disk model can reproduce the full KIN data set reasonably well (Fig.~\ref{fig:null78}), this result seems at odds with the significant K-band excess reported in \citetalias{Absil09}, which would require an equivalent dust surface density 5000 times larger than the solar zodiacal cloud to be reproduced. This is an order of magnitude larger than the level derived to fit the KIN data. This inconsistency forces an examination of dust disk morphologies other than the standard zodiacal disk model, which fails to explain the near- and mid-IR observations simultaneously, and which has no real physical grounding in the case of Fomalhaut. Given the small excess null detected by the KIN (or even considering its 3-$\sigma$ upper limit of just 0.65\%), it is clear that the grains responsible for the near-IR emission are not contributing much at 10 $\mu$m. We will now look for models consistent with this observational result.


    \subsection{Combined Modeling of Near- and Mid-Infrared Data}



We used the GRaTeR code \citep{Augereau1999a} to compute a grid of models that we compare with our full data set consisting of spectro-photometric and interferometric data gathered in the near and mid-IR (see Table.~\ref{tab:data} and Fig.~\ref{fig:fit}).
GRaTeR calculates models for optically thin disks. It is designed to simulate spectral energy distributions (SEDs), images and interferometric data with parametric grain size and radial distributions, or distributions from dynamical simulations. 
Both the scattered light and the continuum emission of dust grains in thermal equilibrium with a star are computed, using the Mie theory and the Bruggeman effective medium method, depending on the material optical constants \citep{bohrenhuffman}.
Particular care is given to the removal of grains when their temperature exceeds the material sublimation temperature. When multi-material grains are used with various sublimation temperatures $T\dma{sub}$, the first material to sublimate is replaced by porosity (affecting the optical properties with respect to compact grains of the same size). We compute a grid of models where we let the following parameters vary: (1) the geometry of the exozodiacal disk (assuming azimuthal symmetry before sky-plane projection) defined through its surface density profile $\Sigma(r) = \Sigma_0 \sqrt{2} \left((\frac{r}{r_0})\uma{-2\alpha\dma{in}} + (\frac{r}{r_0})\uma{-2\alpha}\right)^{-1/2}$
where $\Sigma_0$ is the density at the peak position $r_0$, $\alpha\dma{in}$ the inner slope (fixed to +10 to mimic a sharp inner edge), $\alpha$ the outer slope, (2) the dust size distribution (parameters $\kappa$  and $a\dma{min}$ of the classical power-law $d n(a) \propto a^{\kappa}d a$ valid for grains from $a\dma{min}$ to $a\dma{max}=1~\textrm{mm}$), 
(3) the dust composition $v\dma{C}/v\dma{Si}$, that assumes mixtures of carbonaceous material (volume fraction $v\dma{C}$) and astronomical silicates ($v\dma{Si}$), 
(4) the total disk mass $M\dma{dust}$ in grains up to 1~mm in radius. 
The range of values used for each parameter is summarized in Tab.\ref{tab:parameters}. This parameter space leads to a wide range of different models that we compare with the data using both a classical $\chi^2$ minimization and the statistical Bayesian method described in \citet{2012A&A...539A..17L}. All parameters are assumed to have uniform prior probabilities, at the exception of the treatment we use to account for the inner sublimation radius $r\dma{sub}(a, \textrm{material})$. In that case, we define prior probabilities in order to eliminate all models for which $r\dma{0} < \textrm{min}(r\dma{sub})$.

%

\subsubsection{Fitting data obtained at all wavelengths}

We first attempt to fit all the near and mid-IR data simultaneously (34 data points, see Tab.~\ref{tab:data}).
The shape of the exozodi spectrum (Fig.~\ref{fig:fit} ) requires very hot grains,  which is achieved when these are small and close to the star. Indeed, the probability curves derived from the fit Bayesian analysis and presented in Fig.~\ref{fig:bayes} (red curves) shows that, {\it when considering each parameter independently}, the most probable models are found for very small grains 
($\amin \leq 0.08~\mu m$ 
and $\kappa \leq -5.3$ with 1-$\sigma$ confidence), 
 confined very close to the sublimation distance ($0.07 \leq r\dma{0} \leq 0.14$ AU, 
$\alpha \leq -5.0$)
\footnote{Formally, these probabilities must read as the \textit{probability of the models knowing the data}. It must be noted that for some parameters, the probability curves tend to peak towards the limits of the parameter space, in a physically unmeaningful manner. Thus the confidence intervals we provide must be taken cautiously. In particular, $\kappa$ and $\alpha_{\rm out}$ should be considered as weakly constrained.}.
The model requires that these high-temperature grains include a large fraction of carbonaceous material ($\frac{v\dma{C}}{v\dma{Si}} \ge 10$), to reduce the ten micron emission feature that submicron-sized silicate grains would produce, and to reach higher temperatures. Indeed, the sublimation temperature of carbon is higher than that of silicates
 \citep[$T\dma{sub}(\textrm{C}) = 2000$~K, $T\dma{sub}(\textrm{Si}) = 1200$~K,][]{2009Icar..201..395K} allowing carbon grains to survive closer to the star. The smallest sublimation distances are obtained for the largest grains, with min($d\dma{sub}(\textrm{C})) = 0.07$ AU, and min($d\dma{sub}(\textrm{Si})) = 0.21$ AU (see sublimation curves in Fig.~\ref{fig:dfth_da_dr} for details). 
 
%
A clear secondary peak in the disk inner edge probability curve reveals that a second family of solutions is of statistical relevance, one that uses grains located further out from the star ($r\dma{0} \sim 0.4$~AU). In fact the model with the smallest overall $\chi^{2}$ (see Tab.~\ref{tab:dustmodels})  is found among this second family of solutions\footnote{The most probable solution for individual parameters can indeed be distinct from that found in the 6-dimensions parameter space.}, where $\sim 0.2~\mu m$ grains orbiting in the $\sim$0.4 AU region, dominate the emission, with a characteristic temperature of 1600~K. The disk is largely dominated by thermal emission over scattered light at the wavelengths considered here. The result of this fit to the null and spectrophotometric data is  shown in Fig.~\ref{fig:fit} (red curves). The SED is well fitted for a total dust mass of $\sim4\times10^{-10}M\dma{\oplus}$. The flux contributed by grains of different sizes depending on their distance to the star is illustrated in Fig.~\ref{fig:dfth_da_dr} (upper panels). From Fig.~\ref{fig:fit} (red curves), it is clear that this overall best fit model fails at reproducing the rising null depth observed toward the mid-IR ($\lambda \geq 11\mu m$) with the  KIN in 2008. The model predicts instead a monotonically decreasing contrast toward longer wavelengths and overpredicts the shortest wavelengths null depth from 2008, while staying compatible with the 2007 data. 
Overall, it appears impossible to fit {\it all} the data with a single-annulus / single grain population model, since the VLTI detection requires very hot dust close to the sublimation radius, while the apparent rise of the KIN nulls beyond 11 microns is reminiscent of colder dust. The binomial distribution of the best fit inner radius ($r\dma{0}$, Tab.~\ref{tab:dustmodels}) is also suggestive of a two component dust distribution.

In an attempt to explore this two dust component scenario with the GRaTeR code, which is currently limited to a {\it single} dust population (already exploring a large 5-parameter space, see Tab.\ref{tab:parameters}), we now fit separately the data shortward of 11$\mu m$ (hereafter labelled "SHORT") and the data longward of 11$\mu m$ (hereafter labelled "LONG"). An improved version of the  GRaTeR code, including self consistent modeling and radiative transfer calculations through multiple dust components is currently under development (Lebreton in prep.) but beyond the scope of this paper. 

\subsubsection{Fitting data subsets}

The shortest wavelengths ($\lambda < $ 11 $\mu$m, observations (26 data points, see Tab.~\ref{tab:data} ) can be fitted by a family of models comparable to the one previously discussed: the probability curves (Fig.~\ref{fig:bayes}) reveal that very small grains located very close to the sublimation limit are favored. The solution now has $r\dma{0}$ in the range 0.08-0.11 AU, \textit{i.e.} located as close as possible to the star, next to the size- and composition-dependent sublimation radius. The best fit model is presented in Tab.~\ref{tab:dustmodels} and Fig.~\ref{fig:fit} (blue curves). 
Fig.~\ref{fig:dfth_da_dr} (middle panels) reveals that this solution also consists of sub-micron grains with $a\dma{min}$ in the range 0.01-0.21 $\mu m$. {\it Thus both the near-IR and $\lambda <11\mu m$ data can be fit by the same hot dust ring located at very small radii}. 

Now fitting only the longer wavelengths observations ($\lambda > 11\mu m$, 11 points including Spitzer/IRS upper limits, see Tab.~\ref{tab:data}), significantly different results are found. As can be seen in Fig.~\ref{fig:fit} (green curves) and Fig.~\ref{fig:dfth_da_dr} (lower panels), the KIN observations are well reproduced if grains are located in a ring located further out from the star, and two families of solutions emerge with $r\dma{0}~\sim~0.45$AU or $r\dma{0}~\sim~1.0$AU. 
With an unconstrained $\kappa$, and $a\dma{min}$ close to the radiation pressure blowout size ($a\dma{blow}(\textrm{C}) = 3.5\mu m$, $a\dma{blow}(\textrm{Si}) = 2.3\mu m $), the size distribution appears compatible with the dynamical and collisional constraints. 
Any carbon to silicates ratio is equally probable. 
A wide range of disk masses is allowed, from $10^{-9}~M\dma{\oplus}$ for the  first family of solutions, to $10^{-6}~M\dma{\oplus}$. 
The density profile does not necessarily need to be very steep, a -1.0 power-law index for instance yields a satisfactory fit with a mass of $4\times10^{-7}~M\dma{\oplus}$. While it fails to reproduce the observed K-band excess satisfactorily (Fig.~\ref{fig:fit} SED green curve), this model clearly provides the best fit to the overall KIN data.

Simulating the emission of a fully self-consistent two-rings structure based on a combination  of the "SHORT" and  "LONG" wavelength models presented here is beyond the scope of this work. Yet, the rising slope of the nulls depths measurements produced by the LONG model gives us confidence that a two-peak model is the best solution to reconcile all of the data.


%

\section{Summary}

The near and mid-IR interferometric data for Fomalhaut are consistent with a hot debris disk residing interior to the habitable zone of the Fomalhaut system (which extends from $\simeq$ 4 AU to $\simeq$ 6 AU given the star's luminosity). The description of this hot debris disk as a single population of grains is not sufficient to explain both the near-IR excess flux found by VLTI/VINCI, and the small mid-IR flux reported here.  A possible explanation is that the exozodi has a more complex geometry, for instance a double-peaked structure. 

In order to be consistent with the KIN measurements, the VLTI  near-IR excess requires a large population of small ($\simeq$ 10 to 300 nm) hot dust located very close to the rim defined by the sublimation distance of carbon. Such grains are however much smaller than the radiation pressure blowout size and they should be placed on hyperbolic orbits and ejected very quickly from this region. A similar paradox was found by \citet{Defrere11} for the disk of Vega, and unveiling the origin of this hot dust is still a challenge to debris disk science. The main difficulty is to explain how such high levels of sub-micron grains can be present, when these grains should be blown out by radiation pressure on very short timescales, typically of the order of a dynamical timescale, i.e., much less than a year in these inner regions. Is it because these grains are prevented from moving out by a braking mechanism, and if yes, which one? gas drag? collisions in a very dense and radially optically thick disk? Or is it because the production rate of these small particles is so high in these regions that, despite their fast removal, a significant amount is always present at a given time? These issues will be explored in a forthcoming paper (Lebreton et al., in preparation). 

On the other hand, the KIN observations detected a small mid-IR excess which appears to increase with wavelength between 8 and 13 microns, which constrains the dust location to lie in a colder region where astronomical silicates could survive. While the VLTI excess only probes the inner hotter component, this second dust population is traced by the KIN data.  With micrometer sizes compatible with a classical-collisional equilibrium, mid-IR emission at the level seen, would require an independent dust population, originating from a much more massive population of planetesimals, comparable to the Solar System Main Belt. The low statistics associated with this warm component do not allow firm conclusions on the parameters of this grain population, but it is a serious hint that a secondary zodiacal belt lies within the field of the interferometer, i.e. inside a few AU. 

Finally, we emphasize that our model has some limitations and should not be considered a definitive explanation to the hot excess of Fomalhaut. First, because each component contributes to both subsets of data, a proper parameterization of the double-population should be used to self-consistently describe the exozodi. 
Second, our treatment of dust sublimation is still very coarse as it eliminates instantaneously the grains when they exceed the sublimation temperature. \citet{2009A&A...506.1199K} show that a proper treatment of time-dependent sublimation physics allows grains hotter than the current $T\dma{sub}$ to survive a certain time before vanishing, depending on their size. Allowing for these transient grains would likely result in putting some large grains inside the current sublimation rim, thus impacting the best-fit models. 
In a future study, we will improve these aspects of our model, hopefully reconciling the results with theoretical predictions. Of course, another possibility is that the disk has suffered important variability in the time interval between VLTI and Keck observing runs. More observations are needed to answer this open question, and more generally constrain hot dust transience around mature stars.

\vspace{0.5cm}
\begin{center}
\bf{Acknowledgments}
\end{center}

The Keck Interferometer was funded by the National Aeronautics and Space Administration (NASA). Part of this work was performed at the Jet Propulsion Laboratory, California Institute of Technology, and at the NASA Exoplanet Science Center (NExSci) , under contract with NASA. The Keck Observatory was made possible through the generous financial support of the W.M. Keck Foundation. 
O. Absil, J.-C. Augereau, J. Lebreton and P. Th\'{e}bault thank the French National Research Agency (ANR, contract ANR-2010 BLAN-0505-01, EXOZODI) for financial support.

\bibliographystyle{apj} 
\bibliography{FomalKIN_v6} 



\newpage

\begin{figure}[h!tbp]\begin{center}   \hspace*{-0.4cm}
  \includegraphics[angle=0,width=0.5\columnwidth,origin=tl]{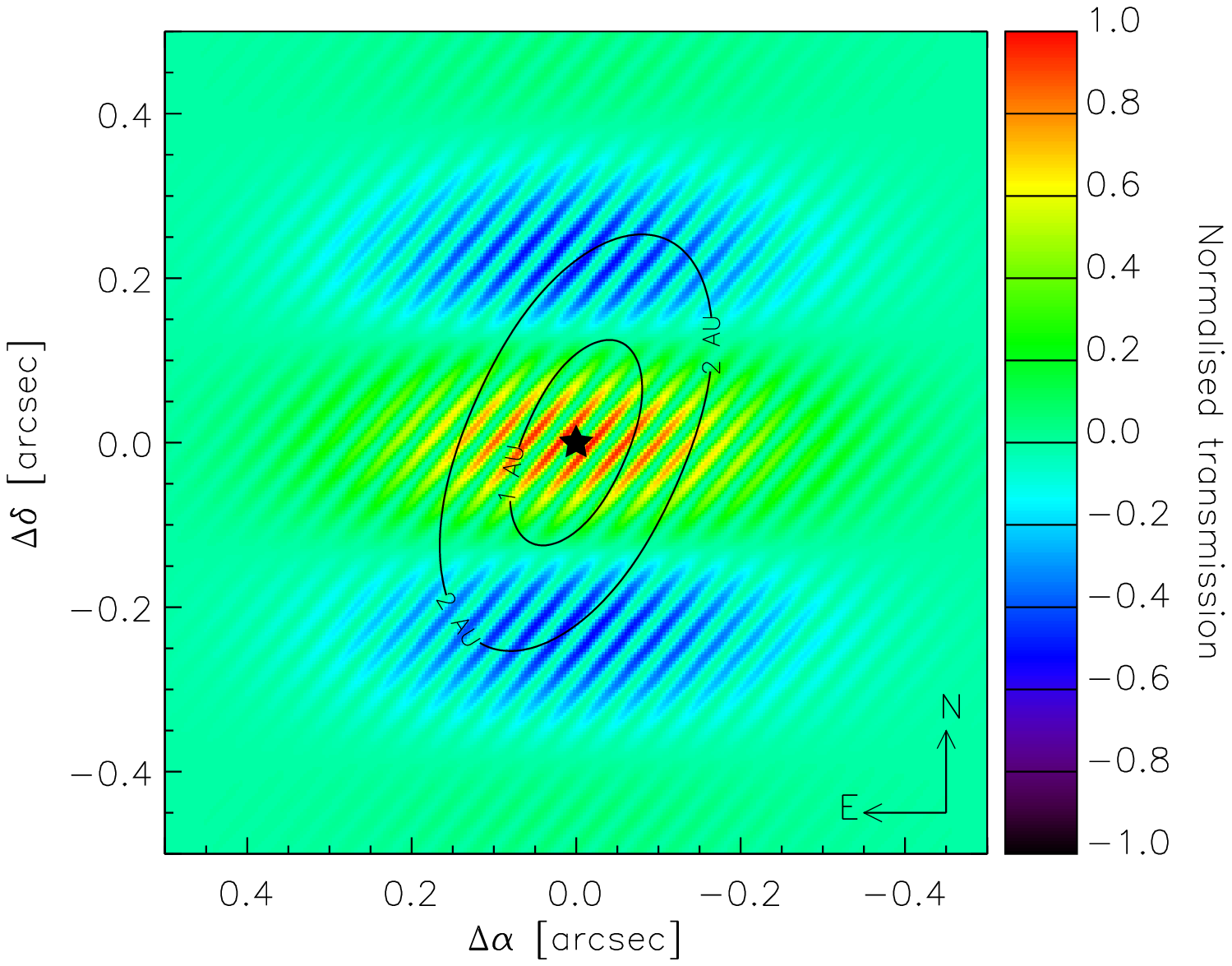}	
  \includegraphics[angle=0,width=0.5\columnwidth,origin=tr]{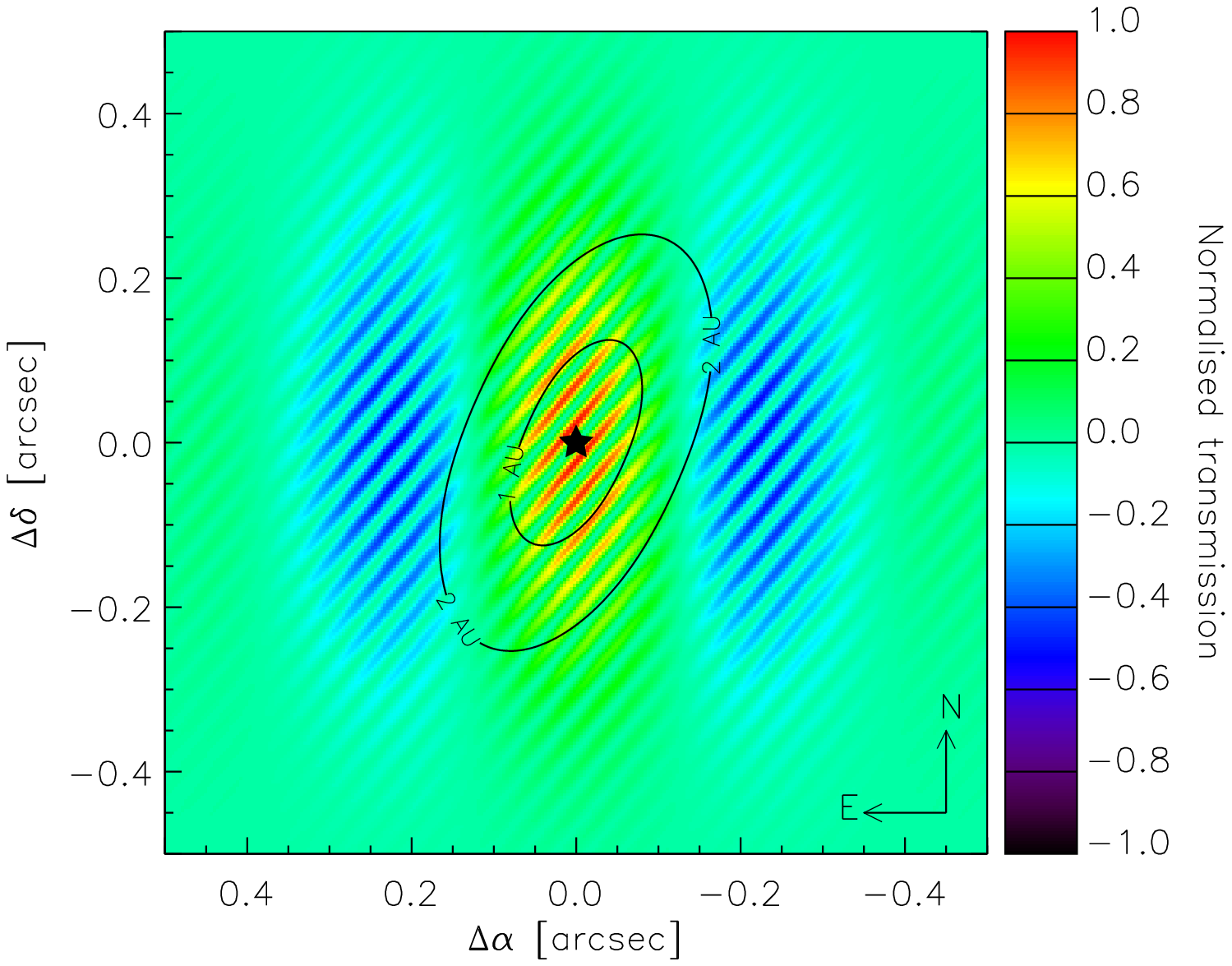}	

    \caption{Left panel: effective Keck Nuller sky transmission at 10\,$\mu$m, when observing Fomalhaut at meridian transit on August 30, 2007 (Julian date: 2454342.95, projected baseline length: 67.6\,m, azimuth: $48\fdeg7$). North is up, East is to the left. High frequency fringes correspond to the long baseline separating the telescopes. The low frequency modulation is produced by interference between the sub-apertures of a single Keck telescope ("cross combiner" fringes); these fringes are aligned with the North-South direction when observing a star at transit. The contours indicate inner regions of the Fomalhaut system ($i=66\deg$, ${\rm PA}=156\deg$), showing that the KIN is sensitive to dust emission in the 0.05 to 2\,AU range. Right panel: Keck Nuller sky transmission at 10\,$\mu$m, when observing Fomalhaut at meridian transit on July 17, 2008 (Julian date: 2454664.07, projected baseline: 67.8\,m, azimuth: $49\fdeg6$). For the 2008 observations, the telescope pupil was rotated by $90\deg$ with respect to 2007. As a result, the low frequency (``cross combiner'' fringes) are now aligned with the East-West direction when observing a star at transit.}\label{fig:tmap12} 
 \end{center}\end{figure}

\newpage

\begin{figure}
\epsscale{.80}
\plotone{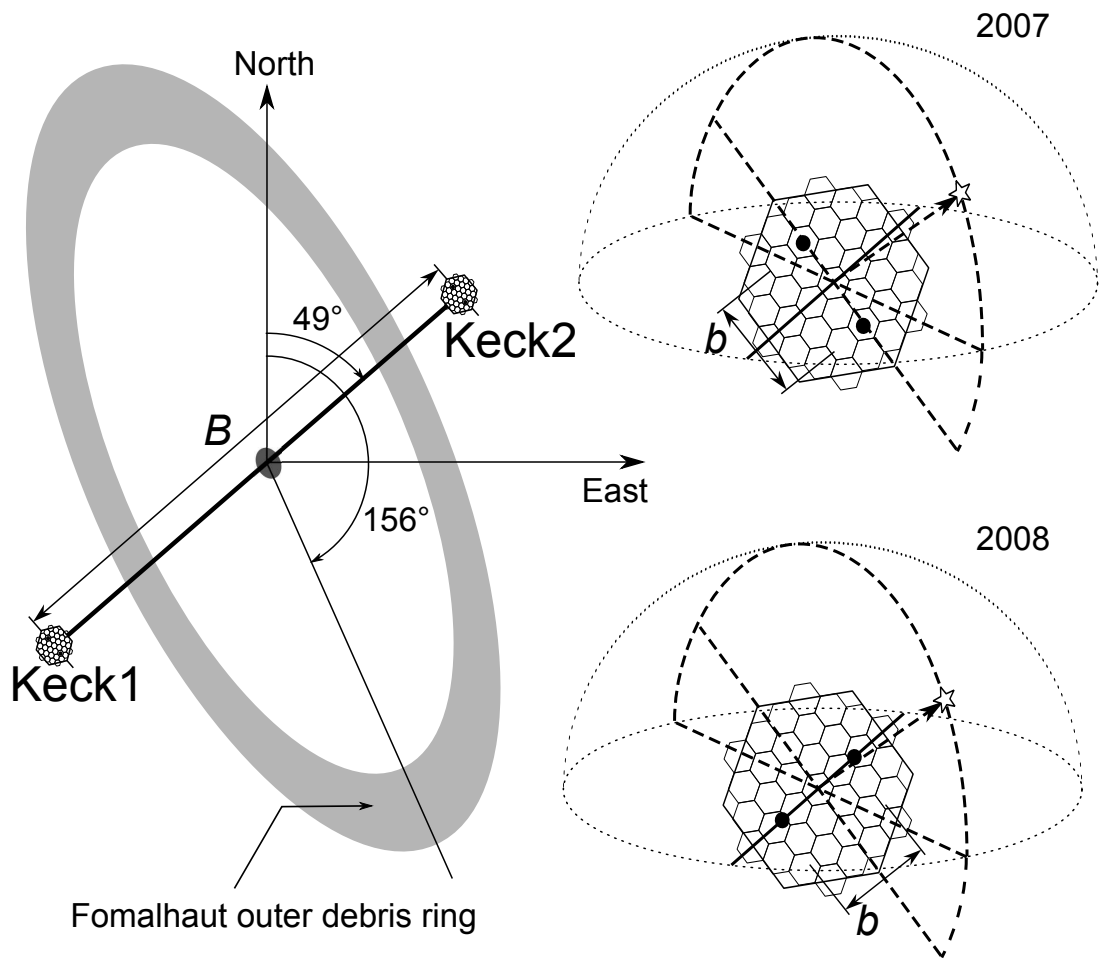}
\caption{Configuration of the long baseline $B$ (left) and of the short baseline $b$ (right) of the Keck Interferoemter Nuller relative to the Fomalhaut planetary system (represented by its cold dust annulus, with a position angle of $156\deg$). All the dashed lines and curves are located within the same plane, including the zenith and the line-of-sight. The two bold dots represent the centers of the two sub-pupils defined on a given Keck pupil. \label{fig:config}}
\end{figure}

\newpage

\vspace{2cm}
\begin{figure}[h!tbp]\begin{center}   \hspace*{-0.4cm}
  \includegraphics[angle=0,width=0.5\columnwidth,origin=tl]{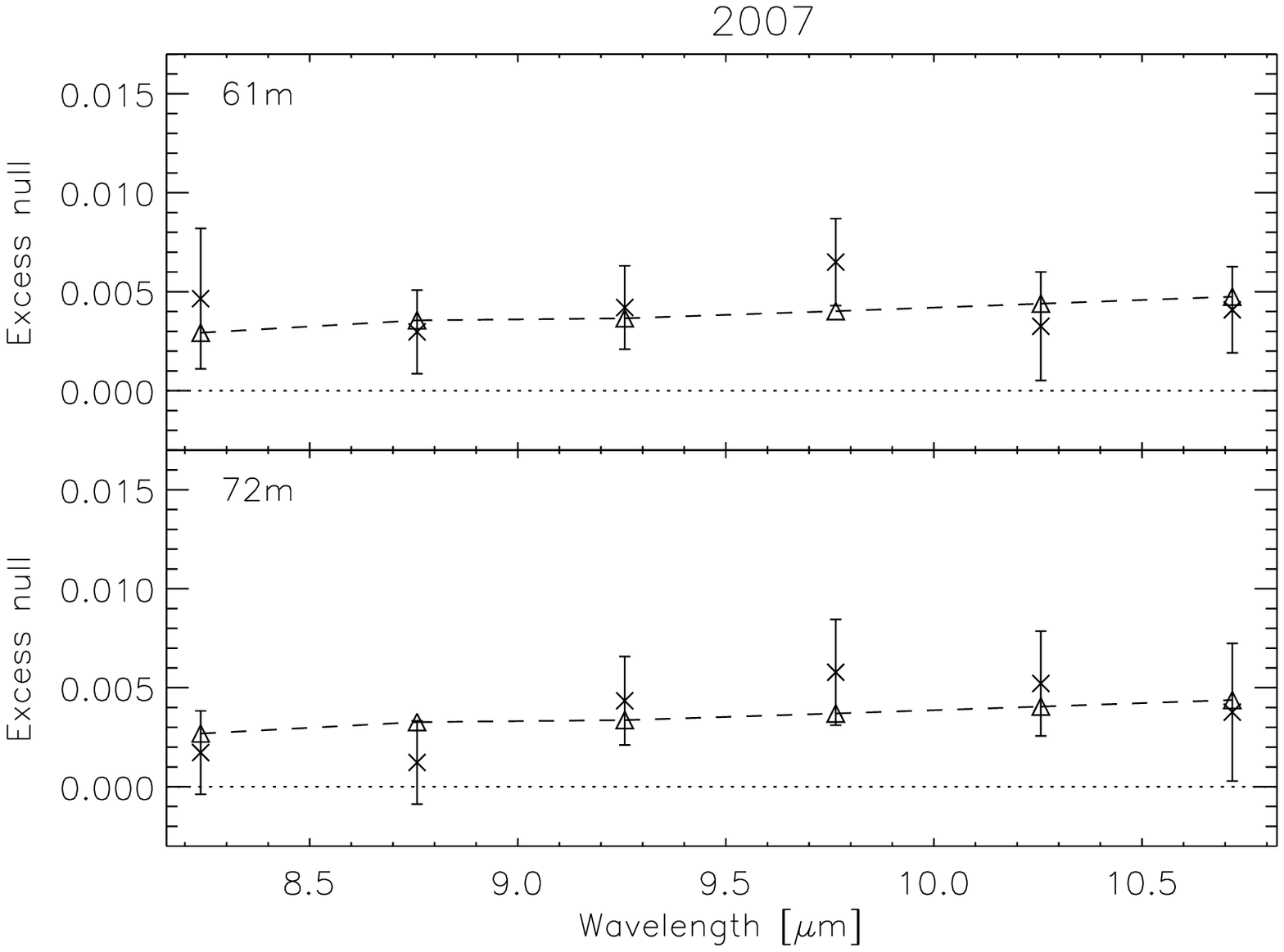}	
  \includegraphics[angle=0,width=0.5\columnwidth,origin=tr]{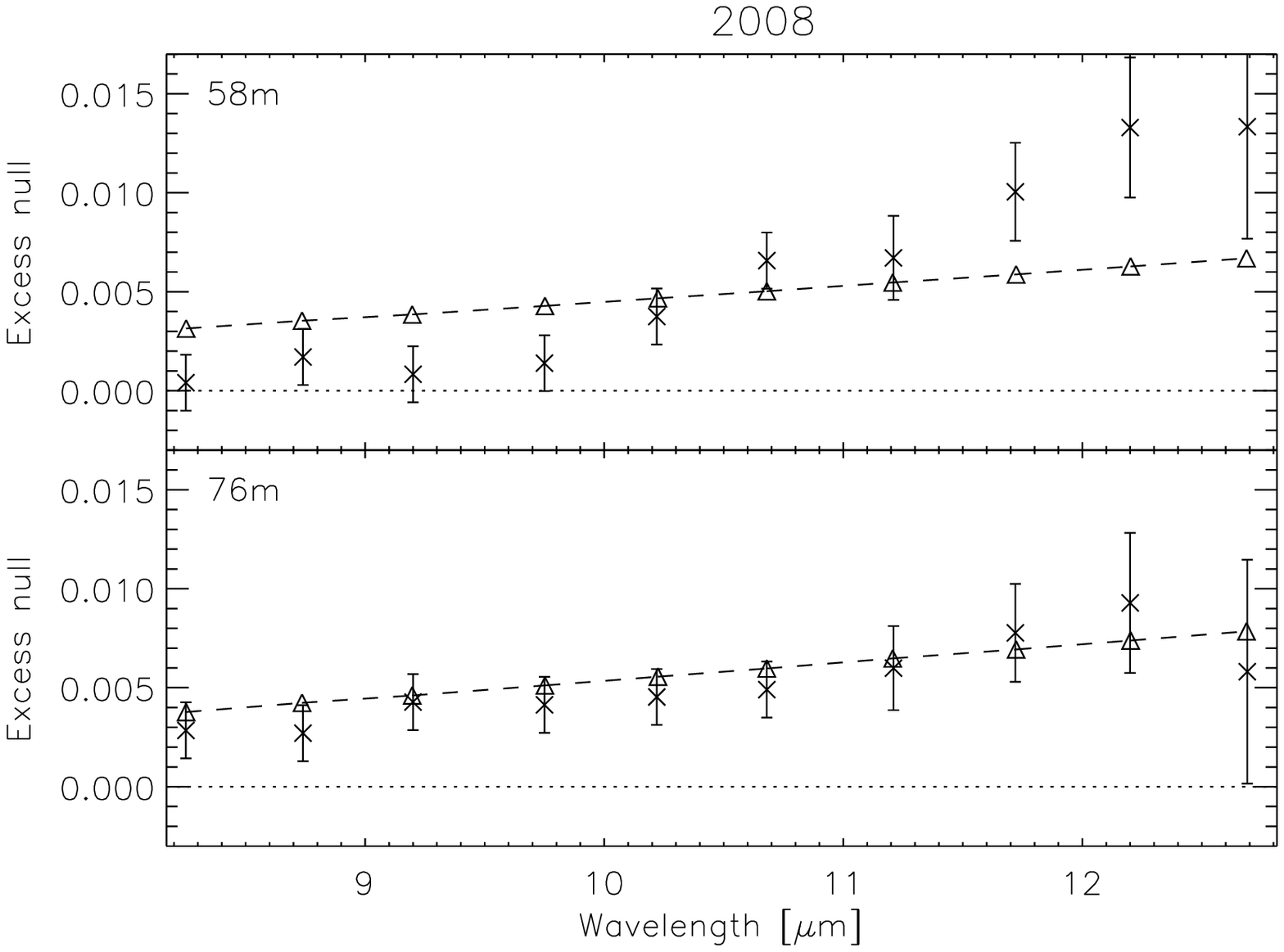}	

 \vspace{4cm}

    \caption{Left panel: 2007 calibrated excess null depth measurements of Fomalhaut plotted as a function of wavelength (crosses with error bars).  Data were obtained at six different baselines and grouped into one short equivalent baseline ($\simeq$ 61\,m) and one long ($\simeq$ 72\,m, see text for details). Right panel: 2008 calibrated excess null depth measurements of Fomalhaut plotted as a function of wavelength (crosses with error bars).  Data were obtained at eight different baselines and grouped into one short equivalent baseline ($\simeq$ 58\,m) and one long ($\simeq$ 76\,m, see text for details).  In both panels, the expected photospheric null depth has been subtracted from the original KIN data to construct the "excess null", which reveals a possible circumstellar excess. The excess null created by a 350-zodi exozodiacal disk is shown with triangles and dashed lines for comparison.}\label{fig:null78} 
 \end{center}\end{figure}

\newpage
%

\begin{figure}[h!tbp]\begin{center}   \hspace*{-0.4cm}
  \includegraphics[angle=0,width=0.43\columnwidth,origin=tl]{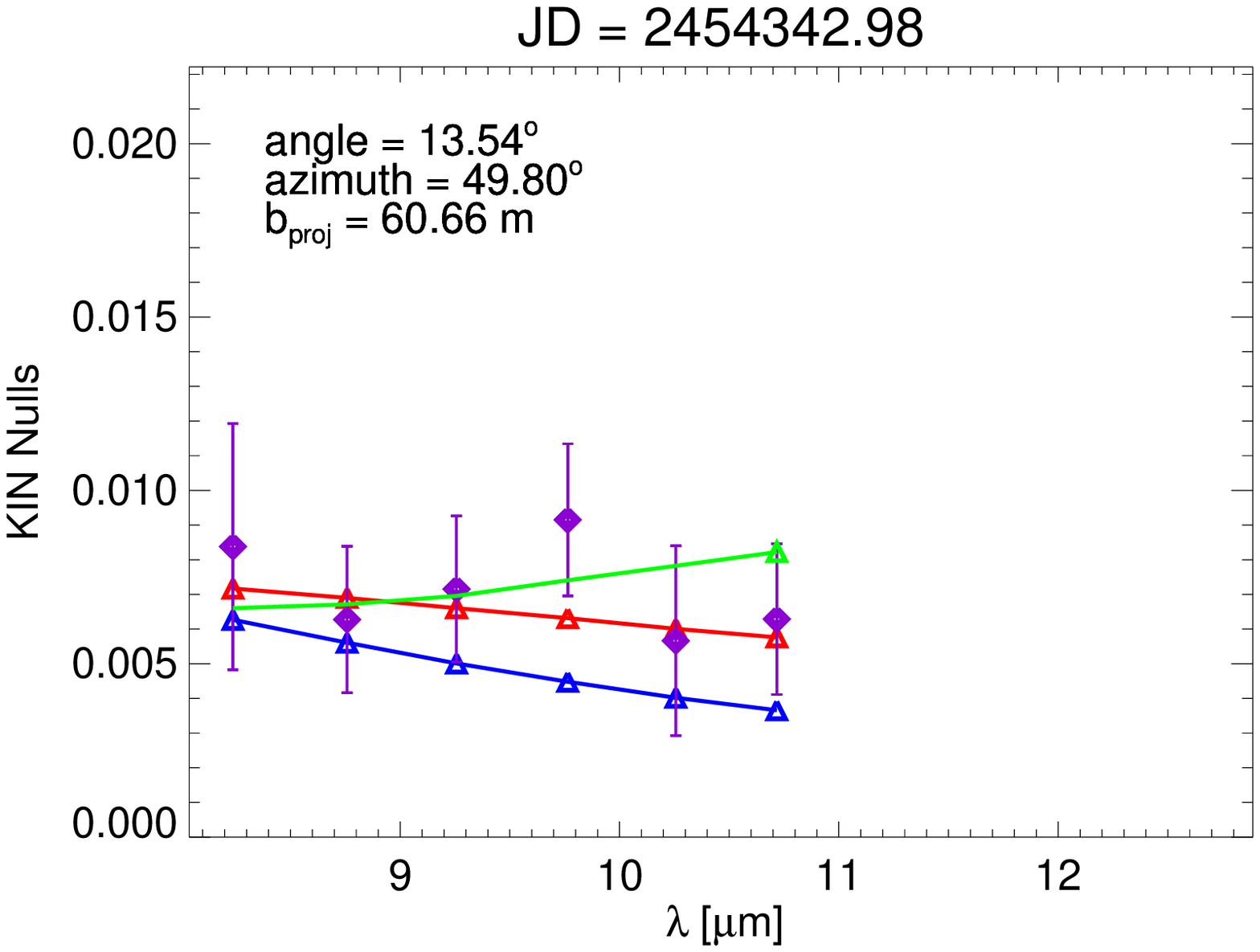}	
  \includegraphics[angle=0,width=0.43\columnwidth,origin=tr]{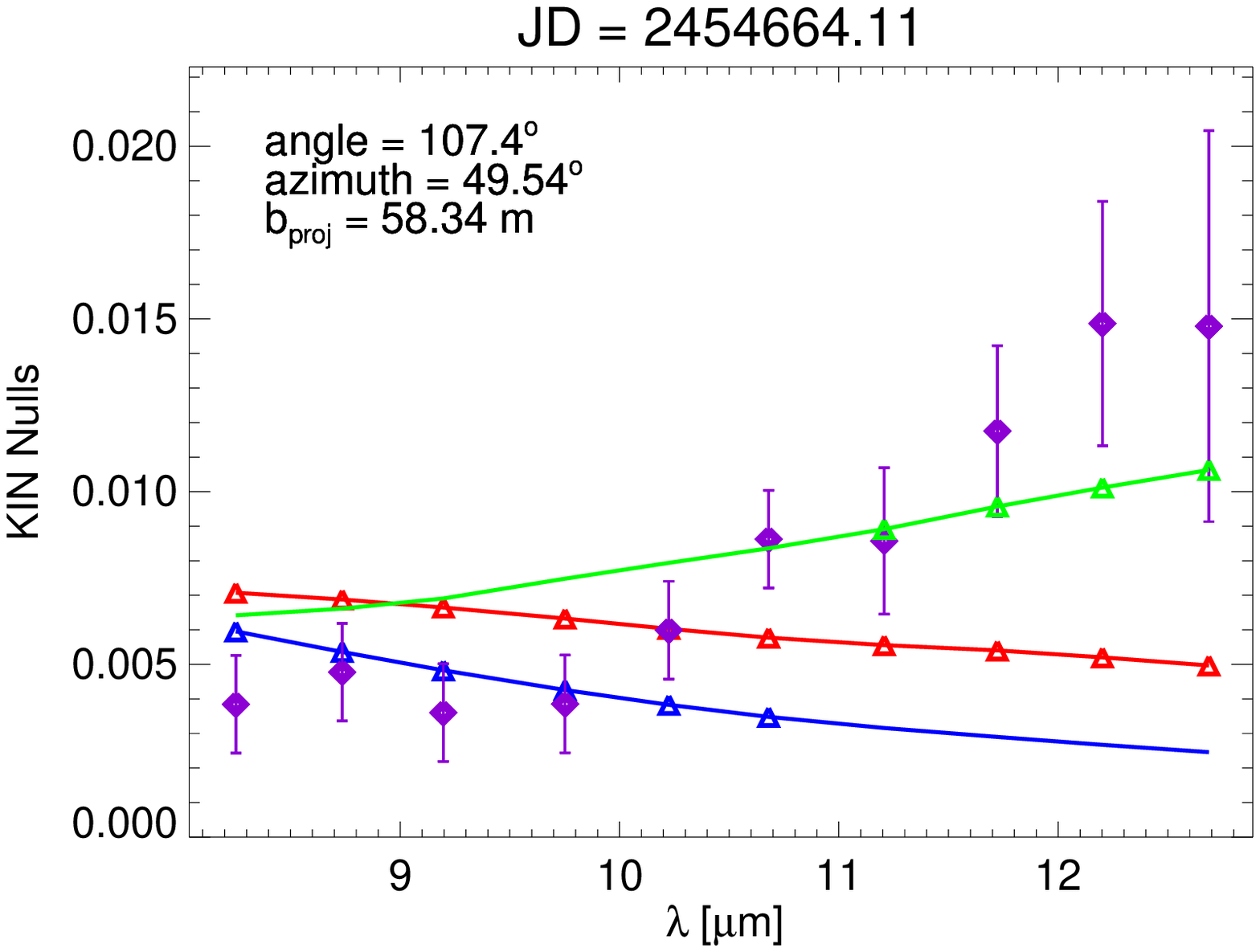}	
  \includegraphics[angle=0,width=0.43\columnwidth,origin=bl]{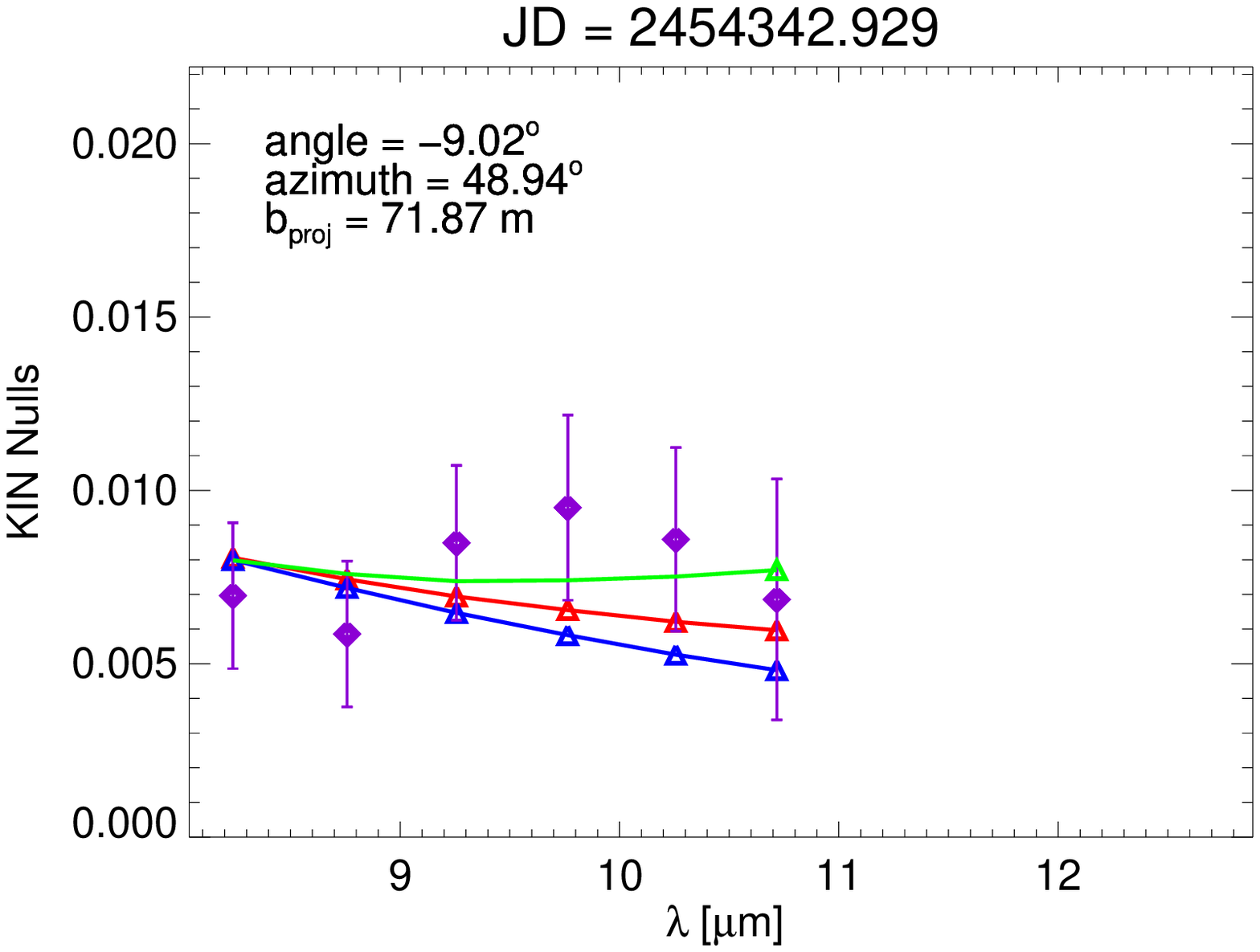}	
  \includegraphics[angle=0,width=0.43\columnwidth,origin=br]{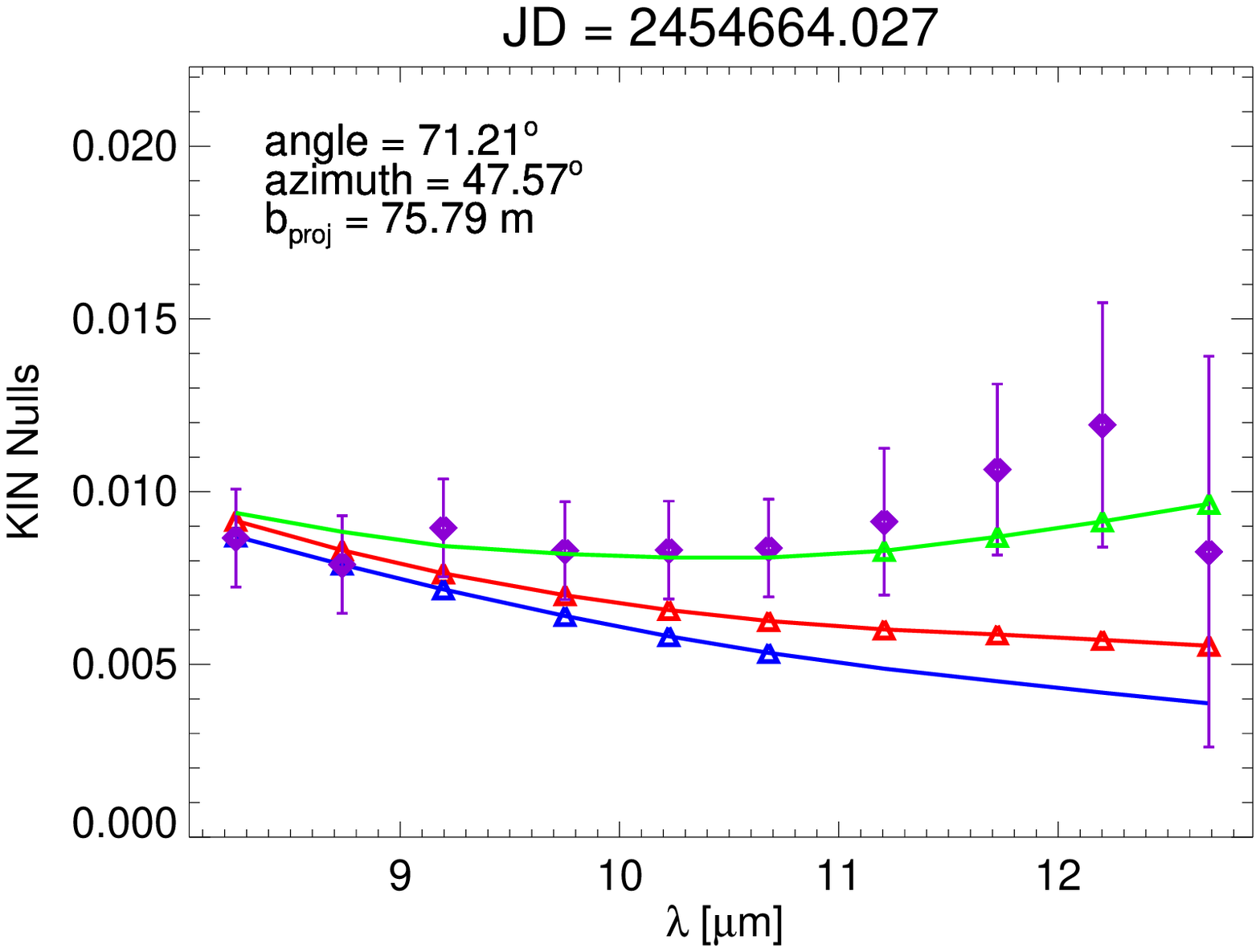}
  \includegraphics[angle=0,width=0.53\columnwidth,origin=bl]{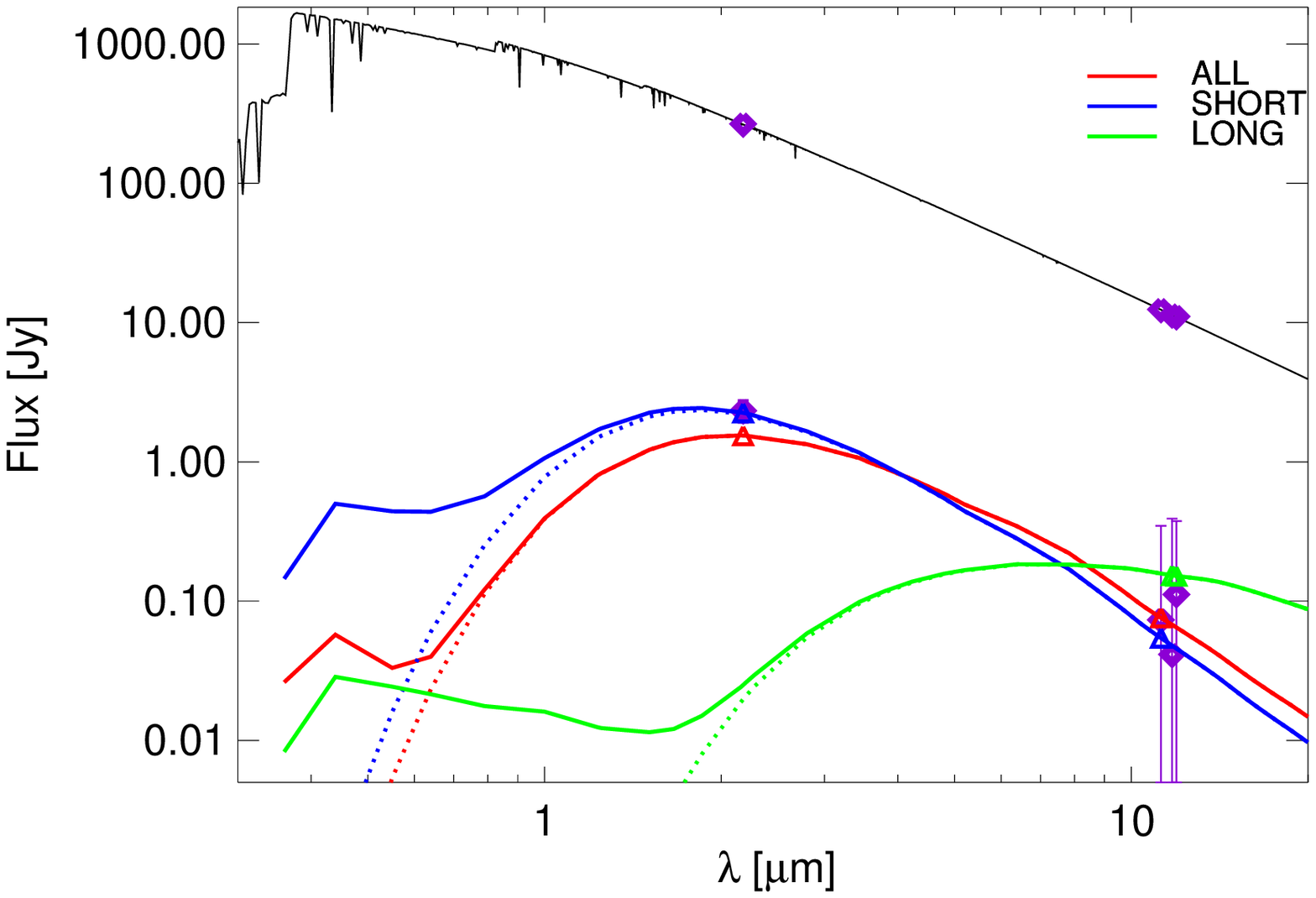}
  
    \caption{Upper 4 panels: KIN calibrated null measurements (purple diamonds) and results of best fit models (Tab.~\ref{tab:dustmodels}, triangles). "ALL" wavelengths model (red curve): uses all interferometric and spectrophotometric data from 2 to 13 $\mu$m.  "SHORT" wavelengths model (blue curve): uses 2 to 11 $\mu$m data only.  "LONG" wavelengths model (green curve):  uses 11 to 13 $\mu$m data only . Bottom panel: solid black line: synthetic photosphere SED model.  Purple diamonds: mid-IR spectro-photometric measurements and K-band VLTI excess. Red, blue and green curves: resulting emission from the disk, for each of the 3 best fit models (same color codes as for upper panels). Solid thick lines: total (scattered + thermal light) emission. Dotted lines: thermal light only. A summary of data points used for each of the 3 models can be found in Tab.~\ref{tab:data}.}\label{fig:fit} 
 \end{center}\end{figure}
 
 \newpage          

\begin{figure}[h!tbp] \begin{center}         \caption{Probability of the models knowing the data calculated over a grid of $\sim~4,000,000$ solutions for the fit to all observations (ALL), the short wavelengths observations ($\lambda \leq 11\mu m$, SHORT), and the long wavelengths observations ($\lambda \geq 11\mu m$, LONG). Depending on the exact composition, the sublimation distances range from approximately 0.07 AU to 0.2 AU for carbon and from approximately 0.2 AU to 0.8 AU for silicates. It is remarkable that the probability curves for the parameter $r_0$ is encompassed within these intervals}\label{fig:bayes} 
   \hspace*{-0.4cm}
    \includegraphics[angle=0,width=1.0\columnwidth,origin=bl]{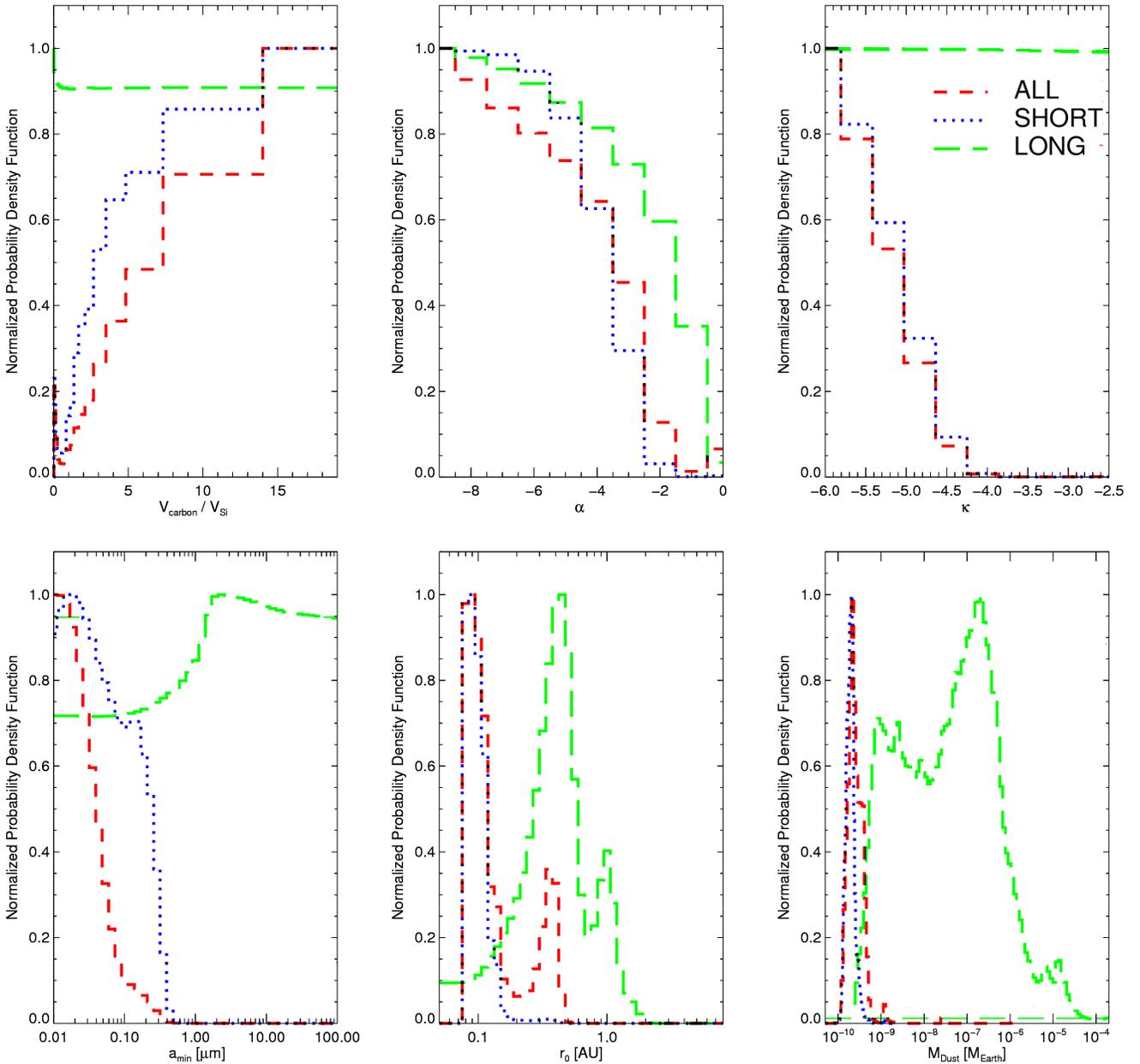}
          \end{center}\end{figure}
 
 \newpage

\begin{figure}[h!tbp] \begin{center}  
\caption{Flux density maps as a function of grain size and distance from the star. Left: $\lambda = 2~\mu m$, right: $\lambda = 13~\mu m$. At each wavelength, maps are shown for 3 models (Tab.~\ref{tab:dustmodels}, either considering  ALL wavelengths data (upper panel), SHORT wavelengths data only ($\lambda \leq 11\mu m$, middle panel), or LONG wavelengths  data only ($\lambda \geq 11\mu m$, bottom panel). For a given model, the 2 and 13 $ ~\mu m$ flux densities essentially differ by a scaling factor, except that larger grains contribute slightly more to the relative emission at $13 ~\mu m$ than at  $2 ~\mu m$ due to their lower temperature. The maps are weighted according to the real (fitted) size distributions and they corresponds to the total flux per unit grain size per elemental annulus of radius distance: the flux in janskys can thus be retrieved through direct integration over distance and grain size. Solid lines: sublimation distances of silicates (green), and carbon (when relevant, white).  The contours are plotted every power of 10 between $10^{-1}$ and $10^{-7}$ of the maximum flux density.}\label{fig:dfth_da_dr}
\hspace*{-0.4cm}
        \includegraphics[angle=0,width=0.47\columnwidth,origin=tl]{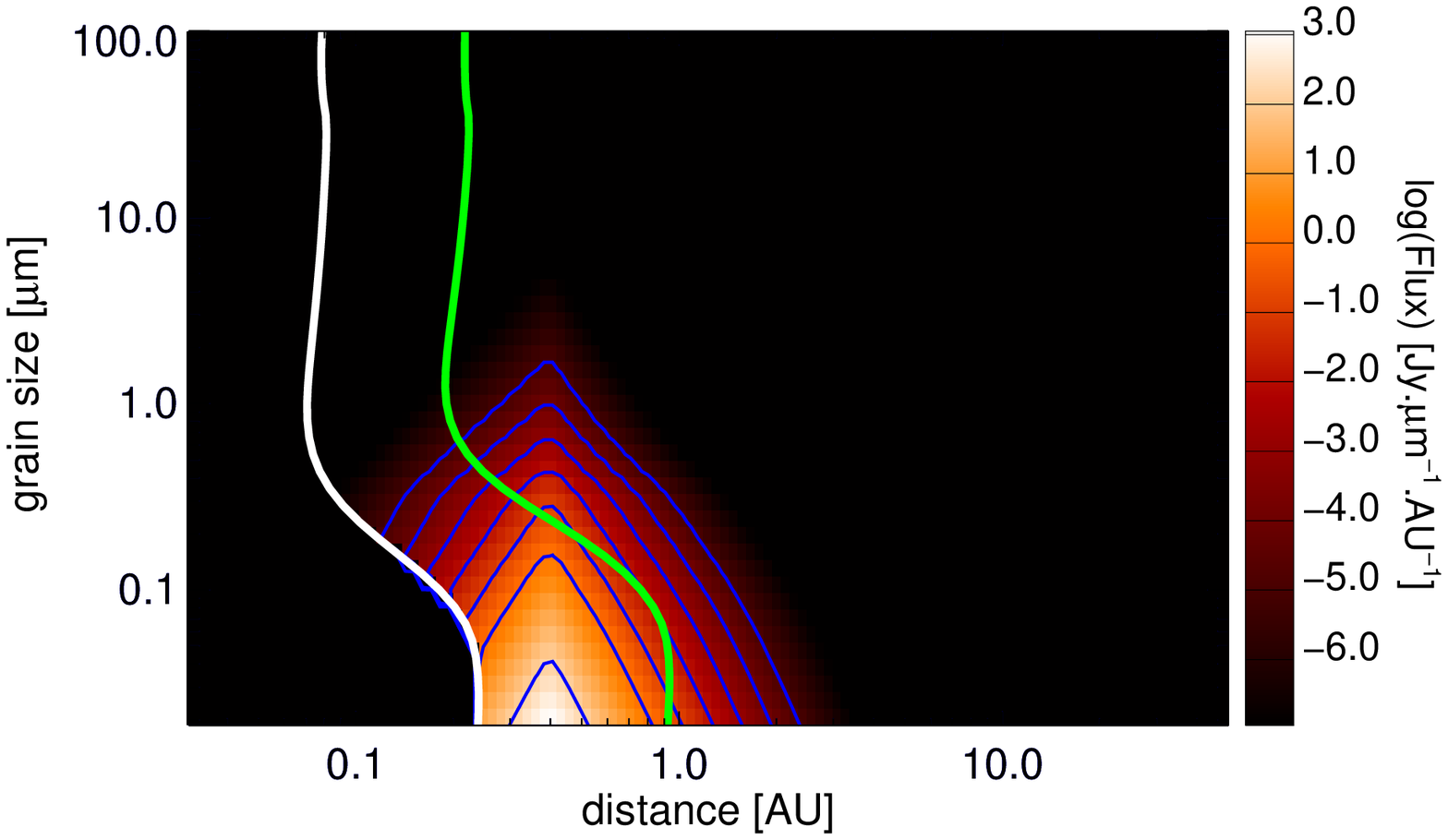}
            \includegraphics[angle=0,width=0.47\columnwidth,origin=tr]{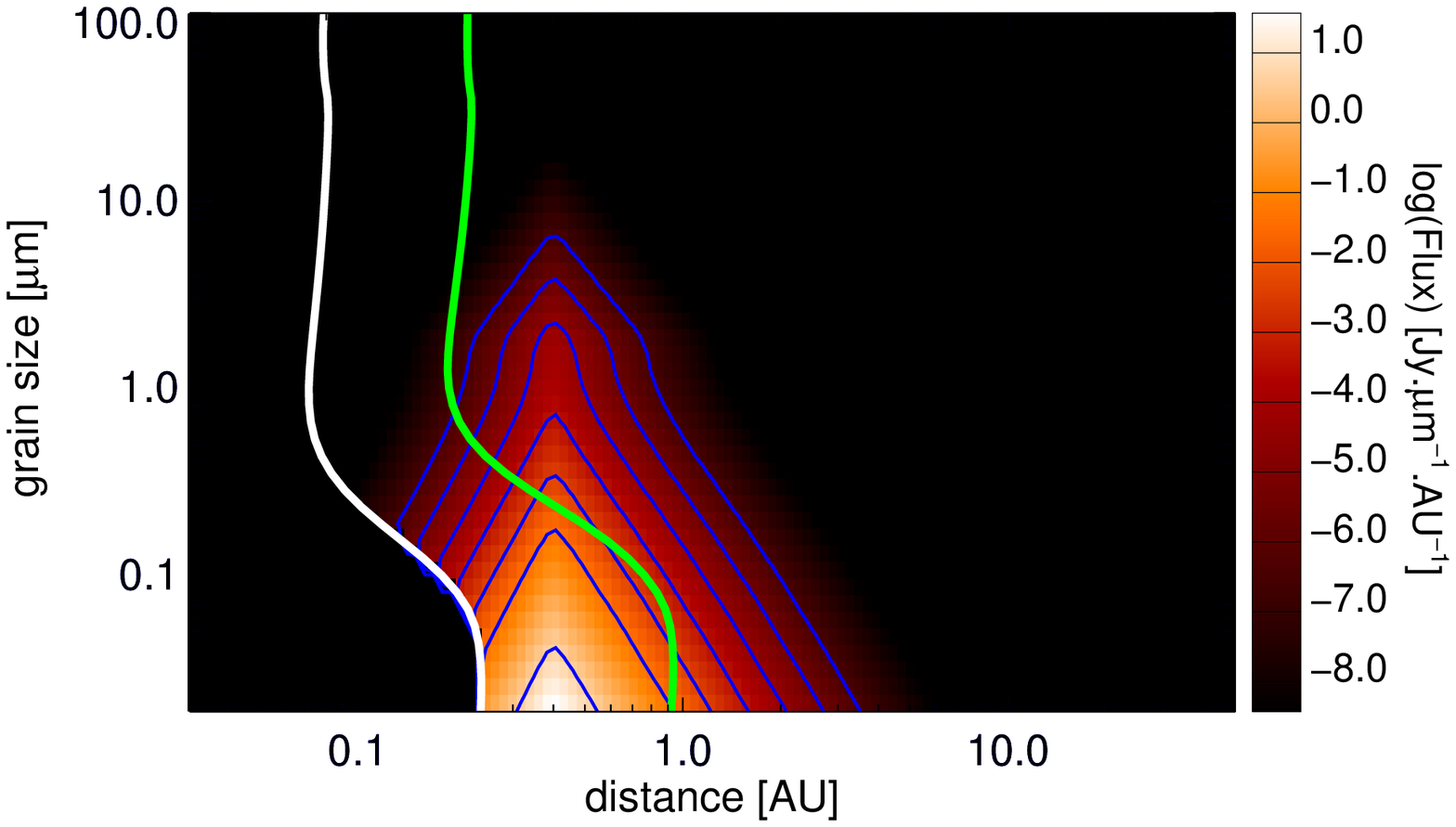}   
        \includegraphics[angle=0,width=0.47\columnwidth,origin=bl]{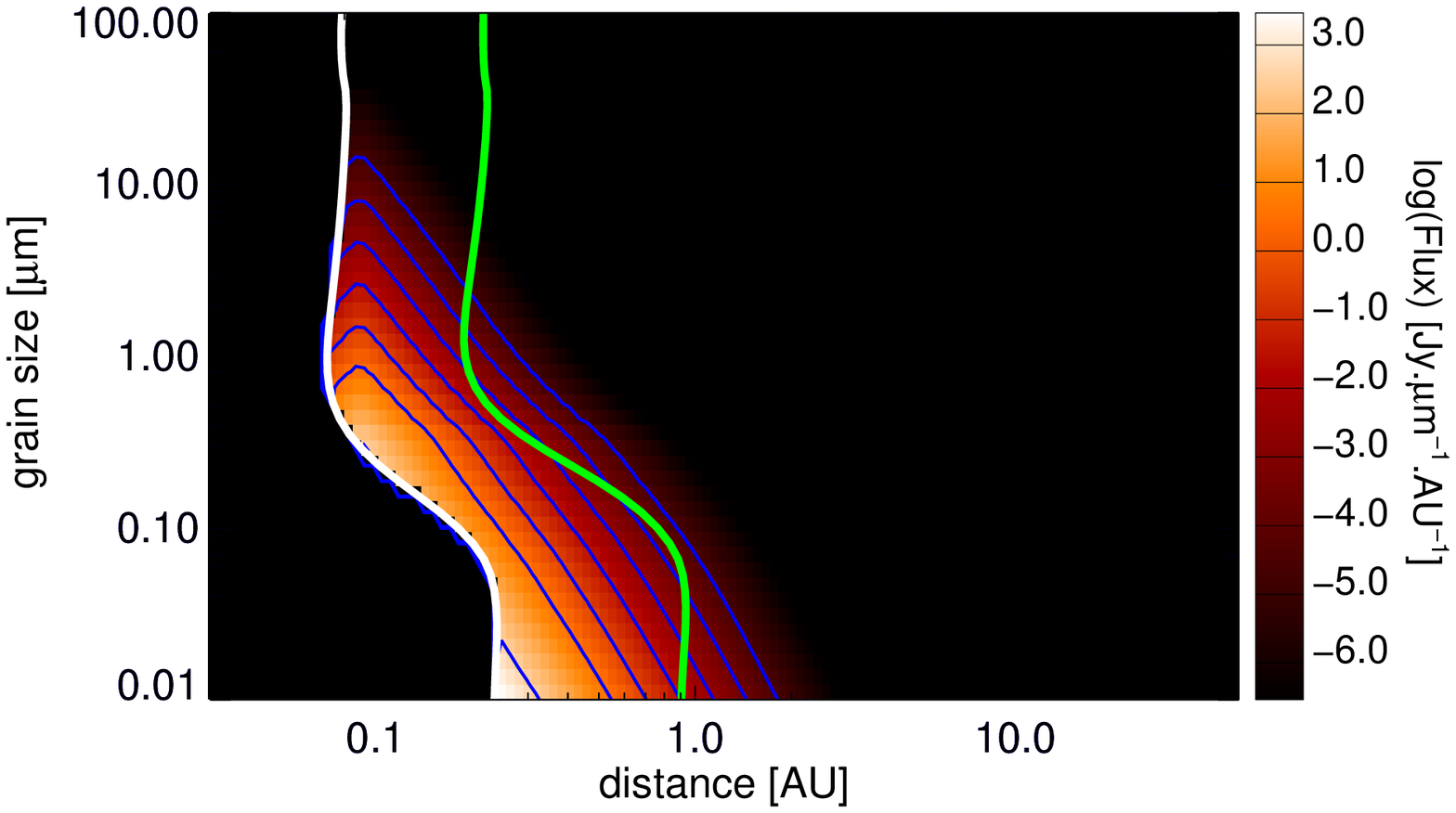}
            \includegraphics[angle=0,width=0.47\columnwidth,origin=br]{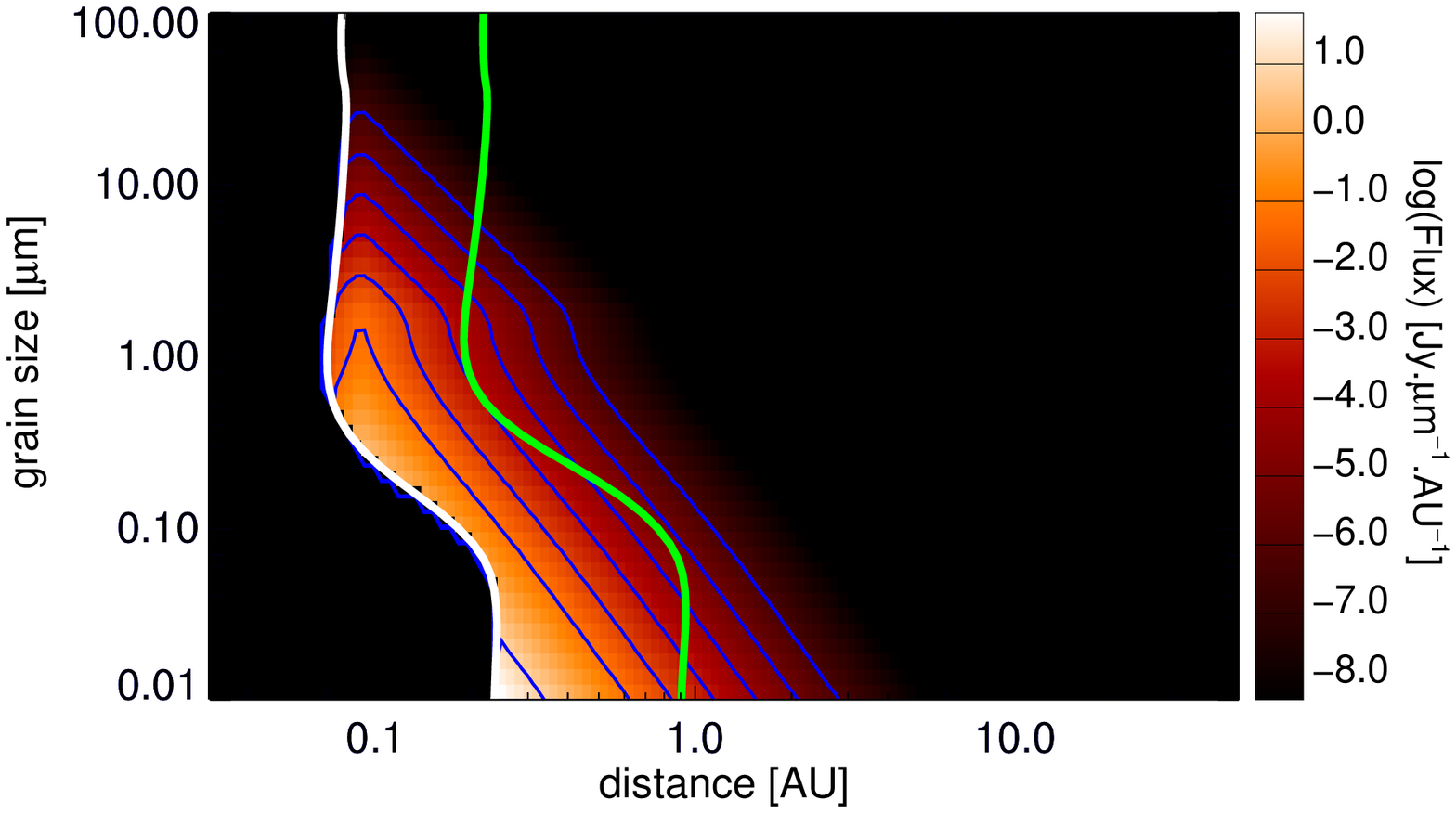}
        \includegraphics[angle=0,width=0.47\columnwidth,origin=bl]{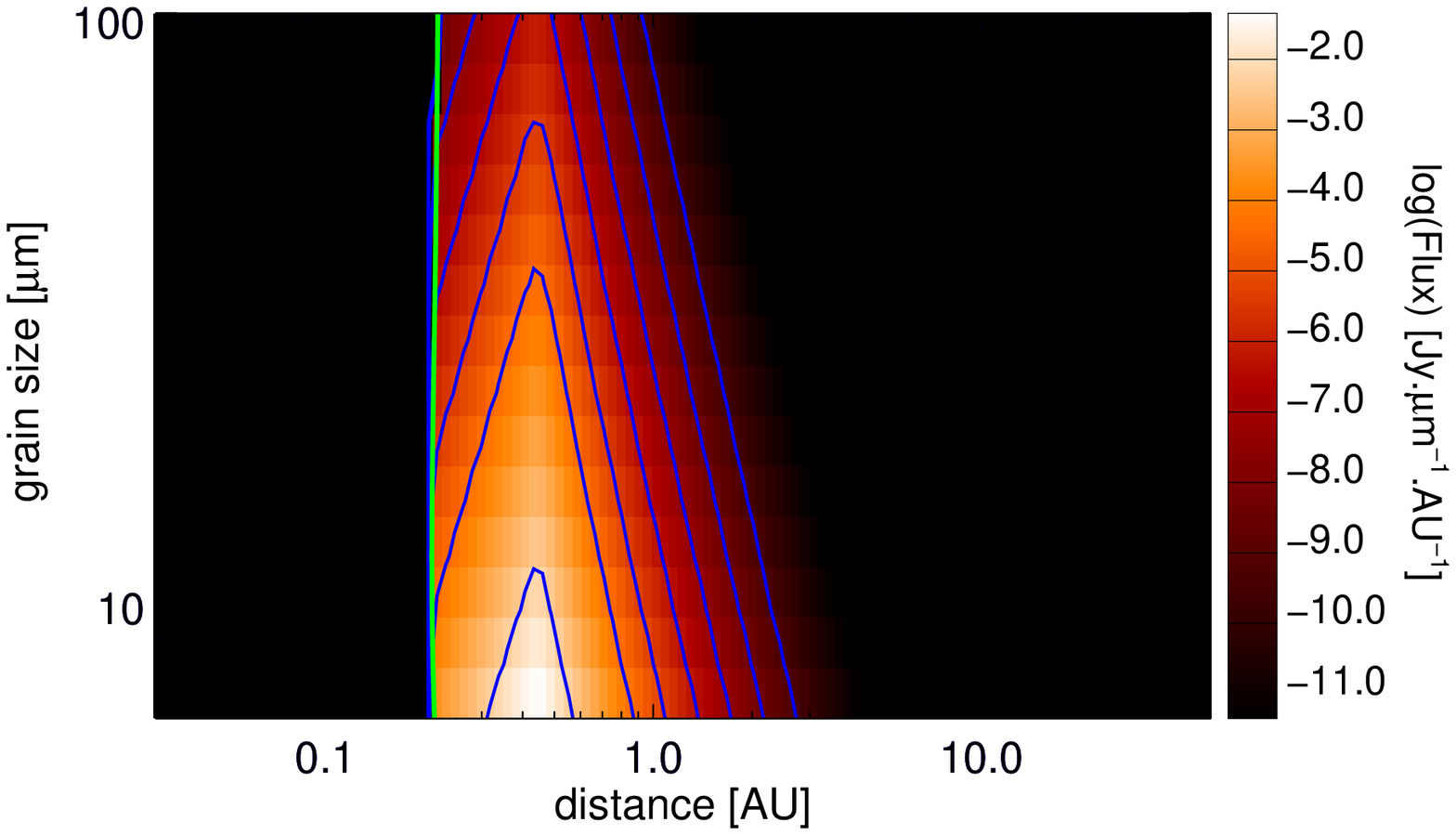}
            \includegraphics[angle=0,width=0.47\columnwidth,origin=br]{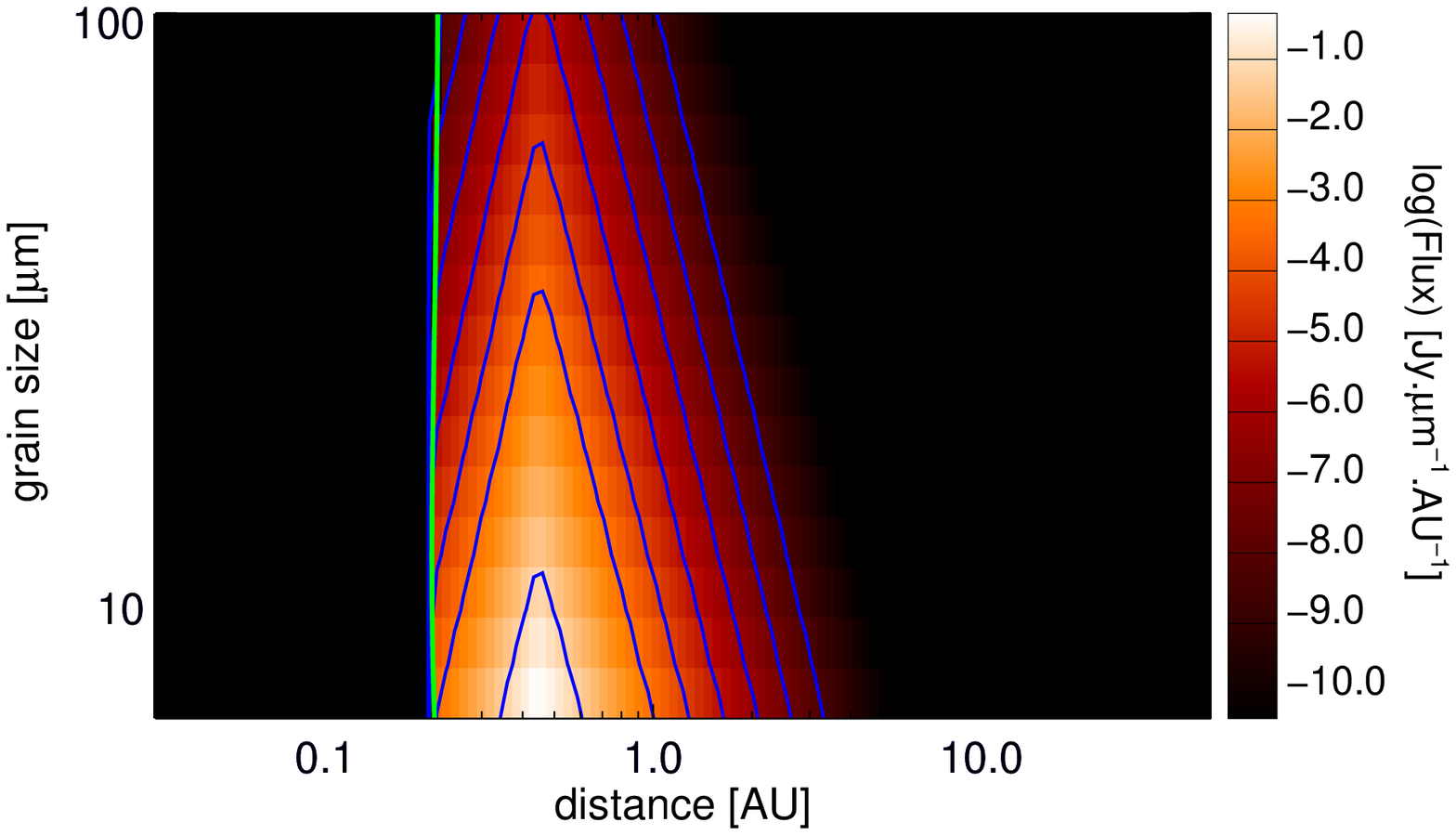}         
          \end{center}\end{figure}


\newpage

\begin{deluxetable}{ccccc}
\tablecolumns{5}
\tablewidth{0pc}
\tablecaption{Calibrators Characteristics \label{tab:cal}}
\tablehead{
\colhead{Star} & \colhead{Type} & \colhead{Dec} & \colhead{R.A.} & \colhead{$\theta_{\rm LD}$ (mas)}
}
\startdata
HD 222547 & K4/5III &  23 41 34 & --18 01 37 & 2.70$\pm 0.16$ \\
HD 210066 & M1III & 22 08 26 & --34 02 38 & 2.58$\pm 0.26$  \\
HD 214966 & M5III & 22 42 22 & --29 21 40 & 3.89$\pm 0.27$ \\
HD 215167 & K3III & 22 43 35 & --18 49 49 & 2.67$\pm 0.16$ \\
\enddata \\
\tablecomments{ Calibrators used for the Fomalhaut observations. }
\end{deluxetable}

\newpage

\begin{deluxetable}{cccccccccccccc}
\tabletypesize{\scriptsize}
\rotate
\tablewidth{0pt}
\tablecaption{Summary Of Fomalhaut KIN Observations \label{tab:obs}}
\tablehead{
\colhead{Date} & \colhead{MJD} & \colhead{$B_{p}$} & \colhead{Az} & \colhead{$N_{8.25}$} & \colhead{$N_{8.74}$} & \colhead{$N_{9.20}$} & \colhead{$N_{9.75}$} & \colhead{$N_{10.22}$} & \colhead{$N_{10.68}$} & \colhead{$N_{11.21}$} & \colhead{$N_{11.72}$} & \colhead{$N_{12.20}$} & \colhead{$N_{12.69}$}
}
\startdata
08/28/07 & 54340.47787 & 62.54 & 49.88  & 1452$\pm$207 & 710$\pm$132 & 842$\pm$140 & 1076$\pm$191 & 313$\pm$172 & 805$\pm$233 & &  & & \\
08/30/07 & 54342.40356 & 76.14 & 47.41  &  666$\pm$148 & 540$\pm$143 & 1074$\pm$159 & 1341$\pm$189  & 1187$\pm$202 & 933$\pm$234 & & & & \\
08/30/07 & 54342.42594 & 72.46 & 48.78  &  699$\pm$144& 717$\pm$135 & 874$\pm$98 & 879$\pm$176 & 898$\pm$134  & 1051$\pm$234 & & & & \\
08/30/07 & 54342.45259 & 67.10 & 49.75  &  758$\pm$164&  623$\pm$156&  621$\pm$145& 649$\pm$178 & 505$\pm$201  & 88$\pm$264 & & & & \\
08/30/07 & 54342.47787 & 61.10 & 49.81  &  459$\pm$199 &  499$\pm$123 & 600$\pm$111 & 948$\pm$181 & 473$\pm$139  & 501$\pm$233 & & & & \\
08/30/07 & 54342.48829 & 58.59 & 49.54  & 631$\pm$129 & 689$\pm$158 &  718$\pm$123& 733$\pm$133 & 915$\pm$198  & 601$\pm$204 & & & & \\
07/17/08 & 54663.51324 & 77.81 & 46.50  & 442$\pm$110 & 462$\pm$118 & 481$\pm$90& 353$\pm$165 & 503$\pm$119  &675$\pm$139 & 406$\pm$190 & 959$\pm$265 & 922$\pm$41 & 351$\pm$681 \\
07/17/08 & 54663.53845 & 73.94 & 48.31  &390$\pm$96  &417$\pm$96 &653$\pm$76 &479$\pm$106  &650$\pm$128  & 511$\pm$147& 639$\pm$156&944$\pm$294 &820$\pm$624 & 322$\pm$998\\
07/17/08 & 54663.59463 & 62.22 & 49.86  &  107$\pm$110 & 281$\pm$98 & 148$\pm$121 & 117$\pm$172  & 365$\pm$139  & 673$\pm$158 & 662$\pm$232 & 902$\pm$261 & 971$\pm$474 & 913$\pm$793 \\
07/17/08 & 54663.62064 & 55.59 & 48.99  & 526$\pm$180 & 250$\pm$151 & 203$\pm$145 & 318$\pm$174  & 618$\pm$164  &  584$\pm$219 & 790$\pm$286 &1630$\pm$309 & 1998$\pm$772 & 1771$\pm$892 \\
07/18/08 & 54664.50661 & 78.33 & 46.17  &  498$\pm$150& 502$\pm$126 & 683$\pm$120& 680$\pm$164  &756$\pm$152  & 739$\pm$181 &946$\pm$266 &565$\pm$383 &1043$\pm$503 & 1502$\pm$723 \\
07/18/08 & 54664.54028 & 73.14 & 48.57  &  750$\pm$176& 633$\pm$160& 626$\pm$157& 670$\pm$143 &494$\pm$170  &575$\pm$172 &846$\pm$212 & 1027$\pm$518 & 1245$\pm$591& 1506$\pm$790\\
07/18/08 & 54664.59729 & 60.90 & 49.79  & 536$\pm$115 & 527$\pm$85& 367$\pm$99& 495$\pm$115 & 605$\pm$166 &878$\pm$140 & 1049$\pm$171& 1179$\pm$241& 1788$\pm$447& 994$\pm$681 \\
07/18/08 & 54664.61968 & 55.12 & 48.88  & 188$\pm$123 & 285$\pm$109 & 230$\pm$116& 146$\pm$139 & 505$\pm$125  &867$\pm$143 & 390$\pm$212 & 810$\pm$272 & 1481$\pm$606 &2388$\pm$898 \\
\enddata \\
\tablecomments{ Date format: month/day/year. MJD: Modified Julian Day = Julian Day - 2400000.5. $B_{p}$: projected baseline in m. Az: baseline azimuth East of North in degrees. $N_{\lambda}$: calibrated astrophysical null (x $10^{5}$) at wavelength $\lambda$.}
\end{deluxetable}

\begin{table}[!hbtp] \centering
  \caption{\small{Summary of the measurements fitted in this study. The last column indicates which data are used for each run of simulations (A: ALL wavelengths from 2 to 13 $\mu$m, S: SHORT wavelengths only, from 2 to 11 $\mu$m, L: LONG wavelengths only, from 11 to 13 $\mu$m). [1] \citet{c2d}}}\label{tab:data}
\begin{tabular}{ccccc}
\hline
Wavelength ($\mu$m)  & Flux[Jy] & Uncertainty ($1\sigma$) & Instrument & fit\\ 
\hline
11.2328 & 12.42 & 0.27    & Spitzer/IRS (c2d) [1] & {-}{-}L \\
11.7363 & 11.37 & 0.35    & Spitzer/IRS (c2d) [1] & {-}{-}L \\
11.9366 & 11.07 & 0.26    & Spitzer/IRS (c2d) [1] & {-}{-}L \\
\hline
Wavelength ($\mu$m)  & Null & Uncertainty ($1\sigma$) & Instrument & fit\\ 
\hline
8.25 & 0.008381 & 0.003551  & KIN 2007 SB & AS-\\
8.74 & 0.006274 & 0.002116 & KIN 2007 SB & AS-\\
9.20 & 0.007156 & 0.002106 & KIN 2007 SB & AS-\\
9.75 & 0.009150 & 0.002196 & KIN 2007 SB & AS-\\
10.22 &0.005664 & 0.002738 & KIN 2007 SB & AS-\\
10.68 &0.006289 & 0.002175 & KIN 2007 SB & AS-\\
\hline
8.25 & 0.006966 & 0.002105 & KIN 2007 LB & AS-\\
8.74 & 0.005859 & 0.002104 & KIN 2007 LB & AS-\\
9.20 & 0.008484 & 0.002234 & KIN 2007 LB & AS-\\
9.75 & 0.009504 & 0.002671 & KIN 2007 LB & AS-\\
10.22 &0.008586 & 0.002648 & KIN 2007 LB & AS-\\
10.68 &0.006852 & 0.003473 & KIN 2007 LB & AS-\\
\hline
8.25 & 0.003848 & 0.001414 & KIN 2008 SB & AS-\\
8.74 & 0.004776 & 0.001414 & KIN 2008 SB & AS-\\
9.20 & 0.003602 & 0.001414 & KIN 2008 SB & AS-\\
9.75 & 0.003854 & 0.001414 & KIN 2008 SB & AS-\\
10.22 & 0.005990 & 0.001414 & KIN 2008 SB & AS-\\
10.68 & 0.008624 & 0.001414 & KIN 2008 SB & AS-\\
11.21 & 0.008571 & 0.002121 & KIN 2008 SB & A-L\\
11.72 & 0.011752 & 0.002475 & KIN 2008 SB &A-L\\
12.20 & 0.014865 & 0.003536 & KIN 2008 SB &A-L\\
12.69 & 0.014790 & 0.005657 & KIN 2008 SB &A-L\\
\hline
8.25 & 0.008658 & 0.001414 & KIN 2008 LB & AS-\\
8.74 & 0.007888 & 0.001414 & KIN 2008 LB & AS-\\
9.20 & 0.008950 & 0.001414 & KIN 2008 LB & AS-\\
9.75 & 0.008292 & 0.001414 & KIN 2008 LB & AS-\\
10.22 & 0.008310 & 0.001414 & KIN 2008 LB & AS-\\
10.68 & 0.008367 & 0.001414 & KIN 2008 LB & AS-\\
11.21 & 0.009132 & 0.002121 & KIN 2008 LB &A-L\\
11.72 & 0.010642 & 0.002475 & KIN 2008 LB &A-L\\
12.20 & 0.011935 & 0.003535 & KIN 2008 LB &A-L\\
12.69 & 0.008260 & 0.005657 & KIN 2008 LB &A-L\\
\hline
Wavelength ($\mu$m)  & Fractional Excess & Uncertainty ($1\sigma$) & Instrument & fit\\ 
\hline
2.18 & 0.0088 & 0.0012 & VLTI fractional excess & AS-\\
\hline 

\end{tabular}

%
\end{table}
%


\newpage

\begin{table*}[h!tpb]\begin{center}\caption{Parameter space explored with GRaTer}\label{tab:parameters}
\begin{tabular}{cccc}
\hline
 Parameter & Explored range &  Values & Distribution\\ 
\hline
\hline
$r\dma{0}$ [AU] & [0.05, \ldots 8.0] & 45 & log \\
$\alpha$& [-9.0, \ldots, 0.0] & 10 & linear \\
$a\dma{min}$ [$\mu$m] & [0.01, \ldots, 100] & 45 & log \\
$\kappa$ & [-$6.0$, \ldots, -$2.5$] & 10 & linear \\
$\dfrac{v\dma{C}}{v\dma{Si}}$  & [$0, $\ldots, $19]$ & 20 & linear\\
$\alpha\dma{in}$& 10 & fixed & -- \\
$a\dma{max}$ [mm] & $1.0$ & fixed & --  \\
$M\dma{dust}$ [$M_{\oplus}$] & $> 0$ & fitted & -- \\

\hline
\end{tabular}\end{center}
\vspace*{-0.3cm}
{\sc Notes -- } We use the astronomical silicates from of \citet{LiGr97} and the ACAR sample of carbonaceous material from \citet{zubko}. $M\dma{dust}$ is scaled independently when adjusting the surface density $\Sigma\dma{0}$ at the peak position to fit the SED.
\end{table*}

\newpage

\begin{table*}[btp]
  \caption{Best-fitting parameters for the three approaches. The bracketed values correspond to the 1-$\sigma$ confidence intervals derived from our Bayesian analysis (possibly several families of solutions can co-exist), while the individual values are those corresponding to the smallest $\chi^2$ on the grid. Some confidence intervals must be taken cautiously (semi-open intervals) because their margins correspond to the limits of the parameter space explored (see Fig.~\ref{fig:bayes}).}\label{tab:dustmodels}
\begin{tabular}{lccccc}\\
\vspace*{-0.2cm}\\
\hline\hline
 & ALL & SHORT & LONG \\
 \hline
\multirow{2}*{$v\dma{C} / v\dma{Si}$}	&[9.6,19[ & [8.3,19[ & [0,19[\\
										&19		& 19 & 0\\
\vspace{0.1cm}\\
\multirow{2}*{$r\dma{0}$ (AU)}			&$[0.07, 0.14]\bigcup[0.33,0.41]$	&	 [0.08, 0.11] &$[0.21,0.62]\bigcup[0.88,1.08]$\\
										&0.40	&0.09 & 0.45\\
\vspace{0.1cm}\\
\multirow{2}*{$\alpha$}		&]-9.0,-5.0] & ]-9.0,-5.2] & ]-9.0,-4.0] \\
										&-9.0 & -8.0 & -9.0\\
\vspace{0.1cm}\\
\multirow{2}*{$a\dma{min}$($\mu$m)}	&]0.01,0.08] &  ]0.01, 0.21] & [1.5,69.0] \\
										&0.02 & 0.01 & 6.6\\
\vspace{0.1cm}\\
\multirow{2}*{$\kappa$}					&]-6.0,-5.3] & ]-6.0,-5.2] &]-6.0,-3.6] \\
										&-6.0 & -6.0 & -6.0\\
\vspace{0.1cm}\\
\multirow{2}*{$M\dma{dust}$($10^{-10}$M$\dma{\oplus}$)}	&	[1.1,5.6]&	[1.5,2.4] & [2.3,6.1$\times10^{5}$]$\bigcup$[1.7$\times10^{6}$,5.6$\times10^{6}$]\\
										&2.2 & 2.2 & 75.5\\
\vspace{0.1cm}\\
{$\chi\uma{2}$ (dof)}					&42.53 (28) & 29.04 (20) & 1.87 (5)\\
\hline
\end{tabular}
\end{table*}




\end{document}